\documentclass{article}

\usepackage{arxiv}

\usepackage[utf8]{inputenc} 
\usepackage[T1]{fontenc}    
\usepackage{hyperref}       
\usepackage{url}            
\usepackage{booktabs}       
\usepackage{amsfonts}       
\usepackage{nicefrac}       
\usepackage{microtype}      
\usepackage{lipsum}		
\usepackage{graphicx}

\usepackage{natbib}
\usepackage{doi}
\usepackage{amsmath, amssymb}
\usepackage{mathrsfs}
\usepackage{bm}
\usepackage[table]{xcolor}

\usepackage[utf8]{inputenc}
\usepackage{amsmath,amssymb}
\usepackage{mathrsfs}
\usepackage{bm}
\usepackage{graphicx}
\usepackage[table]{xcolor}
\usepackage{booktabs}
\usepackage{graphicx}
\usepackage{multirow}
\usepackage{booktabs}
\usepackage{hyperref}
\hypersetup{
    colorlinks=true,
    allcolors = blue,
}
\usepackage{setspace}
\usepackage{algorithm}
\usepackage{algorithmic}
\usepackage{caption}
\usepackage{subcaption}
\usepackage{arxiv}
\usepackage[nohead, nomarginpar, margin=0.75in, foot=.25in]{}

\title{Joint operation of a fast-charging EV hub with a stand-alone independent battery storage system under fairness considerations}

\date{} 					

\author{ {\hspace{1mm}Diwas Paudel, Luke Wolf, and Tapas K. Das}\\
        Department of Industrial and Management Systems Engineering, \\
	University of South Florida,\\
	Tampa, FL  \\
	\texttt{diwaspaudel@usf.edu, lawolf@usf.edu, das@usf.edu} \\
}

\begin{document}
\maketitle

\begin{abstract}
The need for larger-scale fast-charging electric vehicle (EV) hubs is on the rise due to the growth in EV adoption. Another area of power infrastructure growth is the proliferation of independently operated stand-alone battery storage systems (BSS), which is fueled by improvements and cost reductions in battery technology. Many possible uses of the stand-alone BSS are being explored including participation in the energy and ancillary markets, load balancing for renewable generations, and supporting large-scale load-consuming entities like hospitals. In this paper, we study a novel usage of the stand-alone BSS whereby in addition to participating in the electricity reserve market, it allows an EV hub to use a part of its storage capacity, when profitable. The hub uses the BSS storage capacity for arbitrage consequently reducing its operating cost. We formulate this joint operation as a bi-objective optimization model. We then reformulate it into a second-order cone Nash bargaining problem, the solution of which guarantees fairness to both the hub and the BSS. A sample numerical case study is formulated using actual prices of electricity and simulated data for the reserve market and EV charging demand. The Nash bargaining solution shows that both participants can benefit from the joint operation.
\end{abstract}

\keywords{EV charging \and battery storage system \and bi-objective \and Nash bargain \and Second-order cone problem (SOCP)}

\section{Introduction}
Rapid growth of electric vehicles (EVs) has prompted a critical need for an extensive infrastructure of EV charging facilities \cite{McKinsey}. A substantial portion of these facilities is anticipated to be hubs housing several (often many) fast charging stations. The growth in EV charging infrastructure will play a pivotal role in minimizing range anxiety among EV owners further promoting EV adoption. The existing Tesla Supercharging stations are examples of fast-charging hubs \cite{Tesla}. The charging hubs can source the needed power in multiple ways. A primary source of power is the day-ahead (DA) electricity market where hub operators can commit to buying a certain amount of the needed power for every hour of the following day. The real-time (RT) electricity market serves as one of the secondary sources to supply any additional power requirement beyond the day-ahead commitment. The other secondary source of power can be a battery storage system (BSS), which can be built-in or external to the hub. In addition to serving as a secondary source of power, BSS can allow the hub operators to use the storage capacity for arbitrage (i.e., to store power at cheaper prices and discharge when prices are high). 

With the advancement of battery technology, large-scale stand-alone battery storage systems are becoming commonplace. Such BSS can partake in supporting one or more of the entities including the electric grid (by supplying power for the reserve/ancillary markets for load balancing and frequency regulation), renewable electricity-generating farms, and a variety of large load-consuming entities like hospitals, factories, etc. As the cost of batteries continues to fall, large stand-alone BSS is becoming a more mainstream element of our newly built power infrastructure \cite{NREL}. In the rest of the paper, we use the word hub to refer to the fast-charging EV hub and the word BSS to refer to the stand-alone battery storage system.

EV charging hubs with numerous stations will soon join the ranks of significant load-consuming centers and thus will benefit from working with independently operated stand-alone BSS. In this paper, we consider a joint operation between a fast-charging EV hub and a stand-alone BSS. The hub uses BSS capacity to arbitrage by storing power from the DA and RT markets and discharging power to either charge EVs and/or sell power back to the RT market. The BSS is considered to participate in the reserve market, by submitting reserve up and reserve down bids, and also to allow the hub operator to use a part of its storage capacity. It is considered that the objective of the hub operator is to reduce its cost of power in meeting the EV charging demand for the day by optimally sourcing DA and RT power and deciding on the arbitrage strategy. The BSS, on the other hand, aims to maximize its gross profit by optimally using the storage capacity among its two participating activities. To our knowledge, a study examining joint operation between a fast-charging hub and a stand-alone multipurpose BSS, as explained above, has not been presented in the literature.  We undertake such a study by developing a bi-objective optimization model. We then reformulate the model to obtain a Nash bargaining solution that is fair to both participants. The model is implemented on a sample case study problem built with real-world electric power market data and simulated data for EV charging demand and reserve market operations. The results demonstrate the value of cooperation and thus will help to establish new ventures in cooperative operation between hubs and BSS.  

The rest of this paper is organized as follows. A brief review of the related literature is provided in Section \ref{sec: lit review}. We describe the problem in Sections \ref{sec: prob des}  and provide the mathematical formulation in Section \ref{sec: math model}. The details of our solution approach are provided in Section \ref{sec: solution approach} followed by a numerical case study in Section \ref{sec: numerical study}. Finally, the concluding remarks are provided in Section \ref{sec: conclusions}.

\section{Related literature}\label{sec: lit review}
In this section, we present a discussion of the relevant literature that explores the following aspects of our problem: power management of EV charging hubs and the operation of the stand-alone BSS. Based on the review, we highlight the gap in the literature that our paper aims to address.

The power management of EV charging hubs is a widely studied topic in the open literature. These studies primarily consider using electricity from DA and RT markets, while some also consider the use of in-house BSS, to meet the EV charging demands. The study presented in \cite{zheng2020day} approaches hub power management via a profit-maximizing day-ahead power procurement strategy accounting for RT market volatility and no BSS availability. The strategy is derived using a scenario-based stochastic optimization approach. A distributionally robust DA commitment strategy for hub power management is presented in \cite{paudel2023distributionally}, which in addition to the DA and RT power considers the in-house BSS availability. A system state-dependent real-time power management strategy is obtained in \cite{paudel2023deep} via a deep-reinforcement learning model that considers both DA and RT markets as well as in-house BSS. There exists a significant collection of literature that considers EV charging hubs which are also parking garages where EVs remain parked for a significant length of time while also receiving charge. These papers consider effectively scheduling the charging time of each EV depending on their arrival and departure schedules and market price variations to reduce the cost of charging while also supporting network load balancing \cite{subramanian2019two}.  Some of these papers also consider arbitrage using available power in the parked EVs by participating in a variety of power markets including the ancillary market \cite{deforest2018day} \cite{sarker2015optimal}. To our knowledge, there are no hub power management studies that consider working jointly with independently operated stand-alone battery storage systems, which are becoming commonplace in modern-day power infrastructure. In the following paragraph, we review some of the papers that explore some of the potential services that stand-alone BSS can provide to the power grid. 

The literature on the operation of stand-alone BSS primarily focuses on their participation in the reserve/frequency markets. A robust formulation for the operation of a BSS participating in spinning and reserve markets is presented in \cite{kazemi2017operation}. The BSS generates revenue through accepted bids, received as capacity payments, and also through the deployment of accepted bids at the prevailing real-time market prices. A similar study considering the reserve market participation by stand-alone BSS is also provided in \cite{padmanabhan2019battery} and \cite{xu2017scalable}. A bidding strategy for battery storage systems in the secondary control reserve market is examined in \cite{merten2020bidding}. The study indicates that the feasibility of revenue generation through reserve market participation by the BSS in stand-alone mode is limited. However, it becomes viable when operating jointly with a virtual power plant that includes wind, photovoltaic, and thermal power sources. An economic study conducted in Germany explored bidding on both day-ahead and automatic frequency restoration reserve markets by a stand-alone BSS \cite{nitsch2021economic}. The study shows an improved economic potential for stand-alone BSS in 2030 than it was in 2019. In a similar economic study of BSS \cite{hu2022potential}, the authors indicate that present BSS cost and lifetime make it not suitable for the energy market participation but has significant economic potential in the reserve markets. 

Based on the reviewed papers, contemporary Battery Storage Systems (BSS) demonstrate the ability to offer cost-effective services to electricity markets, especially in the reserve market. However, there's a noticeable gap in studies assessing the synergy between a stand-alone BSS participating in the reserve market and fast-charging EV hubs. Therefore, our study aims to address this gap by investigating the joint operation of BSS with a fast-charging hub. Furthermore, our research introduces a novel pathway for expanding the range of services a BSS can provide through its collaboration with a fast-charging EV hub.

\section{Problem Description}\label{sec: prob des}

\begin{figure}[ht]
    \centering
\includegraphics[width=1\linewidth]{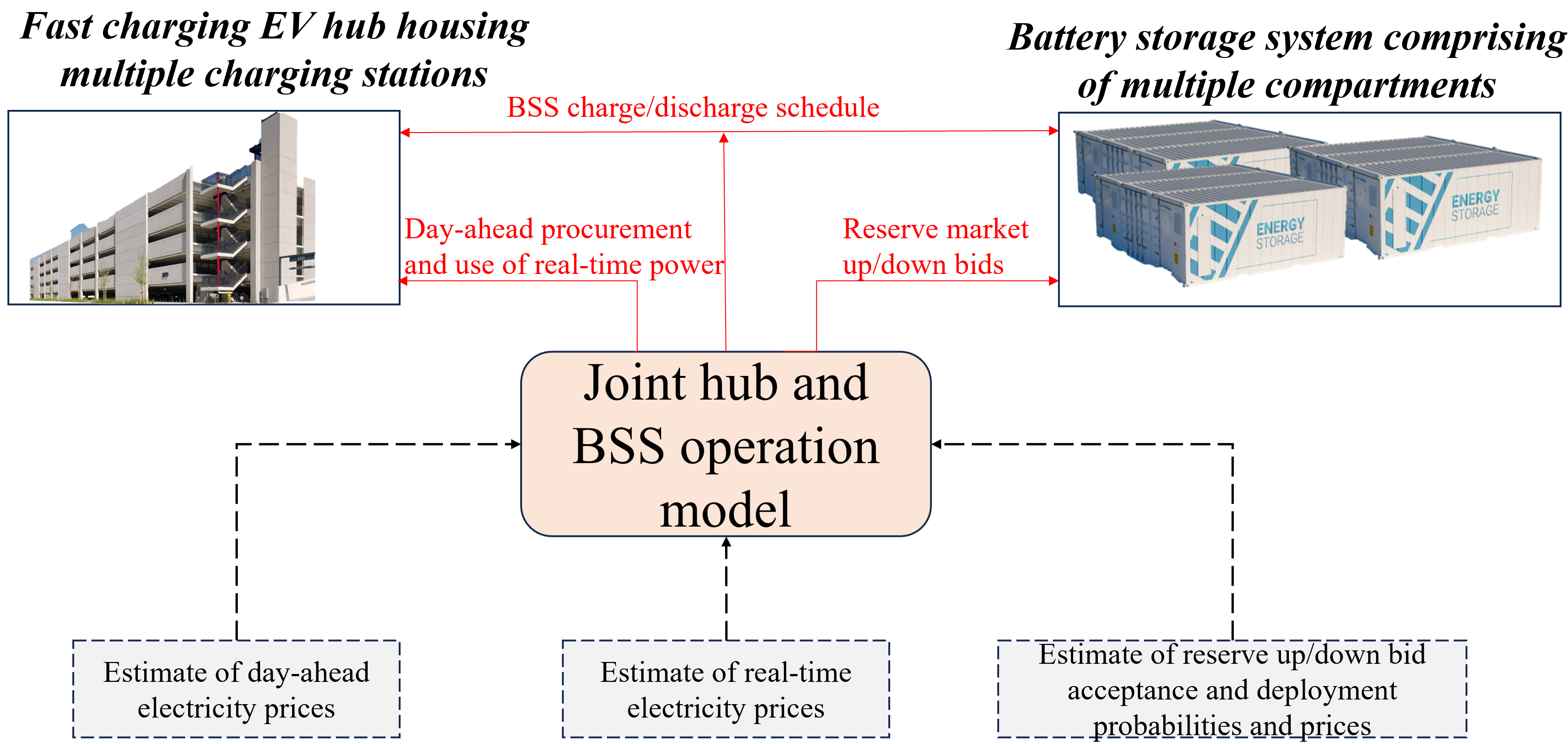}
    \caption{A schematic for joint operation of fast charging EV hub and battery storage system}
    \label{fig: profit_curves}
\end{figure}

We consider a hub with a large number of identical DC-fast charging stations where EVs arrive only for charging and leave promptly after charging.
The hub draws power from the day-ahead and real-time markets of the electricity grid to meet EV charging needs. The primary function of the stand-alone independent BSS is considered to be supporting the reserve market by offering the network operator the use of its storage capacity via reserve up and down services bids. The reserve up capacity represents the amount of stored power that the BSS is willing to supply in real-time in case of an unexpected shortfall in the network. The reserve down capacity, on the other hand, represents the amount of storage capacity that the BSS is willing to offer for storing any excess power in the network. The acceptance of up and/or down bid offers does not imply the actual deployment, i.e., supply or storage of power to/from the grid by the BSS. The deployment occurs sparingly, the probability of which can be accessed from the historical market data. 

The focus of this paper is to study the joint operation of a hub with a BSS that is connected to the same bus of power network for their mutual benefit. In the joint operation mode, the hub uses part of the BSS storage capacity to support its charging needs through arbitrage, i.e. by storing power at a low cost and discharging it when cost-economic, thereby reducing the cost of its operation  The BSS benefits from the additional revenue generated by allowing the hubs to utilize part of its capacity when not deployed for the reserve purposes. The total storage capacity is comprised of multiple compartments each of which can be operated independently by the BSS operator. The key problem addressed in this paper is to determine the optimal joint operational strategies for the hub and the BSS that yield a fair distribution of the added benefits generated from cooperation.

\section{Mathematical model}\label{sec: math model}
In this section, we first present two separate models for the independent operation of the hub and the battery storage system. The model \textbf{P1} is focused on minimizing the hub operator's cost of meeting the charging demand by procuring an optimal amount of power from the day-ahead market for each time interval depending on the expected price volatilities of the real-time market. The model \textbf{P2} maximizes the profit of the BSS operator by optimally selecting up and down bid quantities in the reserve market for each time interval. Thereafter, we present a joint operation model \textbf{P3} where in addition to participating in the reserve market, for each time period, the BSS operator also leases to the hub operator a part of the storage capacity. The hub operator uses the leased capacity for arbitrage and makes a payment for each unit of the stored/discharged quantities. The bi-objective model \textbf{P3} allows both the hub and the BSS operators to improve their objectives through cooperation. 

\subsection{\textbf{P1:} Fast-charging EV hub power management model}
In this model, it is assumed that the hub operator acts independently of BSS and aims to minimize its cost of charging by optimally choosing to use power from the DA and RT markets based on their expected prices. The model determines the optimal hourly quantities for the hub operator to commit in the DA market. 

\begin{subequations}\label{eq: p1}
\begin{equation}\label{eq: p1-power balance da commitment}
    da_t^{com} =  da_{t}^{ev} + da_{t}^{rt}, \quad \forall t \in \mathcal{T}. 
\end{equation}
\begin{equation}\label{eq: p1-upper lim da com}
    da_t^{com} \leq U_t^{com}, \quad \forall t \in \mathcal{T}. 
\end{equation}
\begin{equation}\label{eq: p1-power balance EV load}
    ev_{t}^{l} = da_{t}^{ev} + rt_{t}^{ev}, \quad \forall t \in \mathcal{T}.
\end{equation}
\begin{equation}\label{eq: p1-fcha obj1}
    \min f_{p_1}^a = \min \sum_{t \in \mathcal{T}} \Bigg[ \lambda_{t}^{da} da_{t}^{ev} + (\lambda_{t}^{da} - \lambda_{t}^{rt}) da_{t}^{rt}  + \lambda_{t}^{rt} rt_{t}^{ev}  \Bigg]
\end{equation}
\end{subequations}

Among the constraints of \textbf{P1}, \eqref{eq: p1-power balance da commitment} ensures that the DA commitment $(da_t^{com})$ by the hub operator is fully consumed by charging the EVs in the hub $(da_t^{ev})$ and selling in real-time market $(da_t^{rt})$. The upper bound for the day-ahead commitment quantities is provided by \eqref{eq: p1-upper lim da com}. Similarly, \eqref{eq: p1-power balance EV load} ensures that the EV charging demand in the hub $(ev_t^{l})$ is fulfilled by using the power from the DA commitment and RT market $(rt_t^{ev})$. Note that we consider aggregated EV charging demand for each hub. The goal of the hub operator is to satisfy the total charging demand in the hub in the most cost-effective way as represented by the objective function in \eqref{eq: p1-fcha obj1}, where $\lambda_t^{da}$ and $\lambda_t^{rt}$ are the prices of DA and RT power, respectively. The first term in \eqref{eq: p1-fcha obj1} represents the total cost of using DA power to charge the EVs. The cost of selling DA power in the RT market is captured by the second term. The last term represents the cost of charging the EVs with RT power. 
Problem \textbf{P1} is represented in a compact form as follows. 
\begin{align}
    \begin{split}\label{eq: p1-compact}
        \min_{\bm{x}} f_{p_1}^a(\bm{x}),\\
        \bm{x} \in \bm{X_f}, 
    \end{split}
\end{align}
where, $\bm{x}$ and $\bm{X_f}$ represent the decision variables and the decision space, respectively, of \textbf{P1}. This compact notation is used in the description of the model solution approach in Section \ref{sec: solution approach}.

\subsection{\textbf{P2:} BSS model for reserve market participation}

This model assists an independent BSS operator, participating only in the reserve market, in determining the hourly bids for reserve-up and reserve-down capacities to maximize profit. The reserve-up capacity denotes the power quantity that the BSS is ready to deliver in real-time to address an unforeseen surge in network demand. Conversely, the reserve-down capacity signifies the power amount that the BSS is prepared to absorb (store) in instances of surplus supply within the network. It is considered that the BSS comprises multiple compartments and within each compartment, there is an independent battery unit. These battery units are capable of independent operation, allowing for greater flexibility and control. Each compartment can either charge, discharge, or idle at any time. The total reserve-up and down bids made by the BSS operator are aggregations of compartment bids, which are bounded by the maximum discharge/charge rates of the respective compartments. The up and down bids are assumed to be accepted and deployed with given probabilities, as is considered in \cite{sarker2015optimal} \cite{deforest2018day}.   

\begin{subequations}
\begin{equation}\label{eq: p2-charge or discharge or idle}
    x_{t}^k + y_{t}^k \leq 1, \quad \forall k \in \mathcal{K}, \forall t \in \mathcal{T}. 
\end{equation}
\begin{equation}\label{eq: p2-upper lim up bid}
    \widehat{p}_t^{up, k} <= M^{d, k} y_t^{k}, \quad \forall k \in \mathcal{K}, t \in \mathcal{T}.
\end{equation}
\begin{equation}\label{eq: p2-upper lim down bid}
    \widehat{p}_t^{dn, k} <= M^{c, k} x_t^{k}, \quad \forall k \in \mathcal{K}, t \in \mathcal{T}.
\end{equation}
\begin{equation}\label{eq: p2-deploy up bid}
     \sum_{k \in \mathcal{K}} p_{t}^{up, k} = \pi_{t}^{d, up} \sum_{k \in \mathcal{K}} \widehat{p}_t^{up, k}, \quad \forall t \in \mathcal{T}.
\end{equation}
\begin{equation}\label{eq: p2-deploy down bid}
    \sum_{k \in \mathcal{K}} p_{t}^{dn, k} = \pi_{t}^{d, dn} \sum_{k \in \mathcal{K}} \widehat{p}_t^{dn, k}, \quad \forall t \in \mathcal{T}. 
\end{equation}
\begin{equation}\label{eq: p2-upper lim ch}
    p_{t}^{dn,k} + rt_{t}^{b, k} \leq M^{c,k} x_{t}^k, \quad \forall k \in \mathcal{K}, \forall t \in \mathcal{T}. 
\end{equation}
\begin{equation}\label{eq: p2-upper lim dch}
    p_{t}^{up,k} \leq M^{d,k}y_{t}^k, \quad \forall k \in \mathcal{K}, \forall t \in \mathcal{T}. 
\end{equation}
\begin{equation}\label{eq: p2-upper limit power BSS pack}
     \phi_{t}^{b,k} \leq M^k, \quad \forall k \in \mathcal{K}, \forall t \in \mathcal{T}. 
\end{equation}
\begin{equation}\label{eq: p2-power balance in unleased pack}
    \phi_{t}^{b,k} = \phi_{t-1}^{b,k} + p_{t}^{dn, k} + rt_{t}^{b, k} - p_{t}^{up, k}, \quad \forall k \in \mathcal{K}, \forall t \in \mathcal{T}.
\end{equation}
\begin{equation}\label{eq: p2-bss obj2}
    \max f_{p_2}^b = \max \bigg[  r^{cap} + r^{dep} - c^{\phi} - c^{deg} \bigg]
\end{equation}   
\end{subequations}

Let, $x_t^k$ and $y_t^k$ be the binary variables, where in each time period $t \in \mathcal{T}$, a battery compartment $k \in \mathcal{K}$ can operate in one of the three modes: charging $(x_t^k = 1$ \& $y_t^k = 0)$, discharging $(x_t^k = 0$ \& $y_t^k = 1)$, and idling $(x_t^k = 0$ \& $y_t^k = 0)$  as shown in \eqref{eq: p2-charge or discharge or idle}. Let, $\widehat{p}_t^{up,k}$ and $\widehat{p}_t^{dn,k}$ be the reserve-up and reserve-down bid quantities with corresponding bid prices $\lambda_t^{up}$ and $\lambda_t^{dn}$, respectively. The upper limits for the quantity bids are the maximum allowable charging $M^{c,k}$ and discharging $M^{d,k}$ rates, respectively (see \eqref{eq: p2-upper lim up bid} and \eqref{eq: p2-upper lim down bid}). It is assumed that the up and down bids are accepted, as offered, with probabilities $(\pi_t^{a, up})$ and $(\pi_t^{a, dn})$, respectively. 

The BSS receives a capacity payment for the offered quantity on standby and receives an additional payment if called upon to deploy. The probabilities for up and down deployment are $(\pi_t^{d, up})$ and $(\pi_t^{d, dn})$, respectively. The estimates of the actual dispatch of the up $(p_t^{up, k})$ and down $(p_t^{dn, k})$ capacities are given in \eqref{eq: p2-deploy up bid} and \eqref{eq: p2-deploy down bid}. 
Each compartment can be charged from the actual deployment of the reserve-down capacity and/or with the purchase of power from the RT market $(rt_t^{b,k})$ as given in \eqref{eq: p2-upper lim ch}. The BSS is considered to discharge only to deploy the reserve-up capacities, as in \eqref{eq: p2-upper lim dch}. The maximum amount of power that can be stored in a compartment is controlled by \eqref{eq: p2-upper limit power BSS pack}. The balance of the stored power is provided in \eqref{eq: p2-power balance in unleased pack}, which ensures that the power in compartment $k$ in time $t$, $(\phi_t^{b,k})$, is equal to the power added to $(p_t^{dn,k} + rt_t^{b,k})$  and substracted from $(p_t^{up,k})$ its power level in the previous time period $(\phi_{t-1}^{b,k})$.  

As stated earlier, the BSS participates in the reserve market with the aim to maximize its profit as provided in the objective function \eqref{eq: p2-bss obj2}. The first term in $f_{p_2}^b$ represents the revenue from providing the reserve capacities $(r^{cap})$ in the day-ahead market, of which the components are shown in \eqref{eq: p2-revenue from bid}.
\begin{equation}\label{eq: p2-revenue from bid}
    r^{cap} = \sum_{t \in \mathcal{T}} \Bigg[\lambda_t^{up} \pi_t^{a, up} \sum_{k \in \mathcal{K}} \widehat{p}_t^{up, k}  + \lambda_t^{dn} \pi_t^{a, dn} \sum_{k \in \mathcal{K}} \widehat{p}_t^{dn, k}\Bigg],
\end{equation}
where the first and second terms represent the revenue from accepted reserve up and down bids, respectively, with corresponding prices $\lambda_t^{up}$ and $\lambda_t^{dn}$. 
The second term in $f_{p_2}^b$ represents the revenue from the actual deployment of the reserve capacities $(r^{dep})$, the components of which are shown in \eqref{eq: p2-revenue from deployment}. 
\begin{equation}\label{eq: p2-revenue from deployment}
    r^{depl} = \sum_{t \in \mathcal{T}} \Bigg[ \lambda_t^{rt} \pi_t^{d, up} \sum_{k \in \mathcal{K}} p_t^{up, k} + \lambda_t^{rt} \pi_t^{d, dn} \sum_{k \in \mathcal{K}} p_t^{dn, k}\Bigg],
\end{equation}
where $\lambda_t^{rt}$ is the real-time price of electricity at time $t$.
The cost of purchasing RT power to charge the BSS ($c^\phi$) and the battery degradation cost ($c^{deg}$) in $f_{p_2}^b$ are shown in \eqref{eq: p2-cost of charging BSS} and \eqref{eq: p2-cost of battery degradation}, respectively. The method of calculating the degradation cost is adopted from  \cite{ortega2014optimal}.
\begin{equation}\label{eq: p2-cost of charging BSS}
    c^{\phi} = \sum_{t \in \mathcal{T}} \lambda_t^{rt}\sum_{k \in \mathcal{K}} rt_t^{b, k}.
\end{equation}

\begin{equation}\label{eq: p2-cost of battery degradation}
    c^{deg} = \sum_{t \in \mathcal{T}} \sum_{k \in \mathcal{K}} C^{K} \Big|\frac{m^k}{100}\Big| \Bigg[\frac{\sum_{t \in \mathcal{T}} \big(p_t^{up, k} + p_t^{dn, k}\big)}{BC^k} \Bigg],
\end{equation}
where $m^k$ is the slope of the linear approximation of the battery life as a function of the cycles, $BC^k$ is the capacity, and $C^k$ is the cost of compartment $k \in \mathcal{K}$.

We represent \textbf{P2} in a compact form as \eqref{eq: p2-compact} with $\bm{y}$ being the its decision variables and
$\bm{Y_f}$ being its decision space. 
\begin{align}\label{eq: p2-compact}
    \begin{split}
        \max_{\bm{y}} f_{p_2}^b (\bm{y})\\
        \bm{y} \in \bm{Y_f}
    \end{split}
\end{align}

\subsection{\textbf{P3:} Model for joint operation of fast-charging EV hub and BSS}

We formulate the joint operation of the charging hub and BSS as a bi-objective cooperative optimization model. The joint operation results from the hub's use of a portion of the BSS capacity. The hub operator minimizes the cost of meeting the charging demand by optimally choosing the hourly DA commitment levels and arbitrage strategy for the leased BSS capacity. Whereas, the BSS operator maximizes profit by optimally biding in the DA reserve market and leasing a portion of BSS capacity to the hub operator.

\begin{subequations}
\begin{equation}\label{eq: p3-charge or discharge or idle}
    x_{t}^k + y_{t}^k \leq 1, \quad \forall k \in \mathcal{K}, \forall t \in \mathcal{T}. 
\end{equation}
\begin{equation}\label{eq: p3-upper lim up bid}
    \widehat{p}_t^{up, k} <= M^{d, k} y_t^k, \quad \forall t \in \mathcal{T}.
\end{equation}
\begin{equation}\label{eq: p3-upper lim down bid}
    \widehat{p}_t^{dn, k} <= M^{c, k} x_t^k, \quad \forall t \in \mathcal{T}.
\end{equation}
\begin{equation}\label{eq: p3-deploy up bid}
     \sum_{k \in \mathcal{K}} p_{t}^{up, k} = \pi_{t}^{d, up} \sum_{k \in \mathcal{K}} \widehat{p}_t^{up, k}, \quad \forall t \in \mathcal{T}.
\end{equation}
\begin{equation}\label{eq: p3-deploy down bid}
    \sum_{k \in \mathcal{K}} p_{t}^{dn, k} = \pi_{t}^{d, dn} \sum_{k \in \mathcal{K}} \widehat{p}_t^{dn, k}, \quad \forall t \in \mathcal{T}. 
\end{equation}
\begin{equation}\label{eq: p3-upper limit power BSS pack}
    \phi_{t}^{a,k} + \phi_{t}^{b,k}\leq M^k, \quad \forall k \in \mathcal{K}, \forall t \in \mathcal{T}. 
\end{equation}
\begin{equation}\label{eq: p3-power balance in leased pack}
    \phi_{t}^{a,k} = \phi_{t-1}^{a,k} + da_{t}^{a,k} + rt_{t}^{a,k} - a_{t}^{ev,k} - a_{t}^{rt, k}, \quad \forall k \in \mathcal{K}, \forall t \in \mathcal{T}.
\end{equation}
\begin{equation}\label{eq: p3-power balance in unleased pack}
    \phi_{t}^{b,k} = \phi_{t-1}^{b,k} + p_{t}^{dn, k} + rt_{t}^{b,k} - p_{t}^{up, k}, \quad \forall k \in \mathcal{K}, \forall t \in \mathcal{T}.
\end{equation}
\begin{equation}\label{eq: p3-upper lim ch}
    da_{t}^{a,k} + rt_{t}^{a,k} + p_{t}^{dn, k} + rt_{t}^{b, k}\leq M^{c,k}x_{t}^k, \quad \forall k \in \mathcal{K}, \forall t \in \mathcal{T}. 
\end{equation}
\begin{equation}\label{eq: p3-upper lim dch}
    a_{t}^{ev,k} + a_{t}^{rt,k} + p_{t}^{up, k} \leq M^{d,k}y_{t}^k, \quad \forall k \in \mathcal{K}, \forall t \in \mathcal{T}. 
\end{equation}
\begin{equation}\label{eq: p3-power balance da commitment}
    da_{t}^{com} = \sum_{k \in \mathcal{K}} da_{t}^{a,k} + da_{t}^{ev} + da_{t}^{rt}, \quad \forall t \in \mathcal{T}. 
\end{equation}
\begin{equation}\label{eq: p3-power balance EV load}
    ev_{ t}^{l} = da_{t}^{ev} + \sum_{k \in \mathcal{K}} a_{t}^{ev,k} + rt_t^{ev},  \forall t \in \mathcal{T}.
\end{equation}
\begin{equation}\label{eq: p3-obj1}
    \begin{split}
    \min f_{p_3}^a = & \min \sum_{t \in \mathcal{T}} \Bigg[ \Bigg\{ \lambda_{t}^{da} da_t^{ev}  + (\lambda_t^{da} - \lambda_t^{rt}) da_t^{rt}  + \lambda_t^{rt} rt_t^{ev} \Bigg\} + \\
    & \lambda_t^{da} \sum_{k \in \mathcal{K}} da_t^{a,k} + \lambda_t^{rt} \sum_{k \in \mathcal{K}} rt_t^{a,k} - \lambda_t^{rt} \sum_{k \in \mathcal{K}} a_t^{rt, k}  + c^{deg} \bigg((1 + \alpha) \sum_{k \in \mathcal{K}}\Big(a_t^{ev,k} + a_t^{rt,k}\Big)\bigg)\Bigg].
    \end{split}
\end{equation}
\begin{equation}\label{eq: p3-obj2}
    \begin{split}
    \max f_{p_3}^b = & \max \sum_{t \in \mathcal{T}} \sum_{k \in\mathcal{K}} \Bigg[ \Bigg\{\pi_t^{a,up} \lambda_t^{u} \widehat{p}_t^{up,k} + \pi_t^{a,dn}\lambda_t^d \widehat{p}_t^{dn,k} +\\ & \pi_t^{d, up}\lambda_t^{rt} p_t^{up,k} + \pi_t^{d, dn}\lambda_t^{rt} p_t^{dn,k} - \lambda_t^{rt} rt_t^{b,k}- c^{deg} p_t^{up,k} \Bigg\} + c^{deg}  \alpha\Big(a_{t}^{ev,k} + a_t^{rt,k}\Big)\Bigg].
    \end{split}
\end{equation}
\end{subequations}

The constraints in \eqref{eq: p3-charge or discharge or idle}, \eqref{eq: p3-upper lim up bid}, \eqref{eq: p3-upper lim down bid}, \eqref{eq: p3-deploy up bid}, and \eqref{eq: p3-deploy down bid} are same as in \textbf{P2}. The total power stored in compartment $k$ by the hub $(\phi_t^{a,k})$ and the BSS $(\phi_t^{b,k})$ must be less than or equal to the total storage capacity of the compartment, see \eqref{eq: p3-upper limit power BSS pack}. The balance of power stored by the hub in compartment $k$ is maintained using \eqref{eq: p3-power balance in leased pack}. This ensures that the power stored by the hub ($\phi_t^{a,k}$) at time $t$ is equal to the power in the previous time ($\phi_{t-1}^{a,k}$) plus the power added from DA commitment ($da_t^{a,k}$) and/or the RT market ($rt_t^{a,k}$) minus the power discharged to meet EV charging demand ($a_t^{ev,k}$) and/or the power sold back to the grid ($a_t^{rt,k}$). Similarly, the balance of power stored by the BSS in a compartment is maintained by \eqref{eq: p3-power balance in unleased pack}. The upper limits of the power added to and discharged from a compartment by the hub and BSS are maintained by \eqref{eq: p3-upper lim ch} and \eqref{eq: p3-upper lim dch}, respectively. The balance of hourly usage of DA commitment and fulfillment of EV charging demand are ensured in \eqref{eq: p3-power balance da commitment} and \eqref{eq: p3-power balance EV load}, respectively. The objective functions for the hub and the BSS operators in joint operation mode are provided by \eqref{eq: p3-obj1} and \eqref{eq: p3-obj2}, respectively. The objective function for the hub in \eqref{eq: p3-obj1}, in addition to the terms of the objective function of \textbf{P1} shown within braces, has four other terms: cost of storing DA power, cost of storing RT power, cost of selling BSS power to RT grid, and the cost of leasing battery compartments. The objective function for BSS in \eqref{eq: p3-obj2} contains all the terms, within braces, from the objective function of \textbf{P2}. The only additional term accounts for the profit from leasing BSS capacity to the hub. Equation \eqref{eq: p3-compact} below represents the compact form of \textbf{P3}, where $\bm{z}$ denotes the vector of decision variables and
$\bm{Z_f}$ denotes the decision space. 
\begin{align}\label{eq: p3-compact}
    \begin{split}
        \min_{\bm{z}} f_{p_3}^a,  \max_{\bm{z}} f_{p_3}^b\\
        \bm{z} \in \bm{Z_f}.
    \end{split}
\end{align}

\section{Solution approach}\label{sec: solution approach}



The formulation \textbf{P1} is a linear program and \textbf{P2} is a mixed integer linear (MILP) program, both of which can be solved using any commercial solver. 
Let $d_1$ be the solution of problem \textbf{P1}, i.e., $d_1 = \{\min_{\bm{x}} f_{p_1}^a : \bm{x} \in \bm{X_f}\}$, and $d_2$ be the solution of problem \textbf{P2}, i.e., $d_2 = \{\max_{\bm{y}} f_{p_2}^b : \bm{y} \in \bm{Y_f}\}$. The problem \textbf{P3} is a bi-objective mixed integer linear program and can be solved using algorithms such as $\epsilon$-constraint method \cite{mavrotas2009effective}, the outcome of which would be a Pareto frontier where each point on the frontier is a vector comprising objective function values for both the players i.e., the hub and the BSS. The Pareto frontier with multiple solutions does not provide a rule to select a solution that is convenient to both the players. A Nash bargaining solution (NBS) of \textbf{P3} can identify a single solution from the frontier that is fair to both participants \cite{nash1950bargaining}  \cite{charkhgard2022magic}. This solution satisfies a set of axioms to obtain a fair bargain for all the players. The axioms, as proposed by Nash \cite{nash1950bargaining} include individual rationality, Pareto optimality, symmetry, linear invariance, and independence of irrelevant alternatives. This is the advantage of NBS compared to any other method (e.g., \cite{khanal4658986criterion}) that provides a single Pareto optimal solution. The Nash bargaining problem for \textbf{P3} can be formulated as follows, where $d_1$ and $d_2$ are the disagreement points that represent the payoff of each player under no cooperation. 

\begin{align}\label{eq: Nash social welfare}
    \begin{split}
        \max\quad &(d_1 - f_{p_3}^a)(f_{p_3}^b - d_2)\\
             &\bm{z} \in \bm{Z_f},\\
             &f_{p_3}^{a} \leq d_1,\\ 
             &f_{p_3}^{b} \geq d_2.
    \end{split}
\end{align}
The terms $(d_1 - f_{p_3}^a)$ and $(f_{p_3}^b - d_2)$, in the objective function above, capture the added benefits from cooperation for the cost-minimizing hub operator and the profit-maximizing BSS operator with respect to their disagreement points. The above formulation \eqref{eq: Nash social welfare} is a maximum multiplicative program and is not solvable as it is. Hence we present a second-order cone program (SOCP) reformulation as follows. 

\begin{equation}\label{eq: Nash SOCP}
    \begin{split}
        \max \gamma\\
        0 \leq \gamma \leq \sqrt{\tau_1^0 \tau_2^0}\\
        0 \leq \tau_1^0 = d_1 - f_{p_3}^a\\
        0 \leq \tau_2^0 = f_{p_3}^b - d_2\\
        \bm{z} \in \bm{Z_f}
    \end{split}
\end{equation}
The above formulation is a SOCP since any constraint of the form $\{u, v, w \geq 0: u \leq \sqrt{vw}\}$ is equivalent to $\Big\{u,v, w \geq 0: \sqrt{u^2 + (\frac{(v-w)^2}{2}} \leq \frac{v + w}{2}\Big\}$. The formulation in \eqref{eq: Nash SOCP} can be solved to optimality using most commercial solvers. Interested readers are referred to \cite{ben2001polyhedral} for more details on SOCP reformulation.

\section{Numerical Case Study}\label{sec: numerical study}

\begin{figure}[htp]
    \centering
    \includegraphics[width=0.5\linewidth]{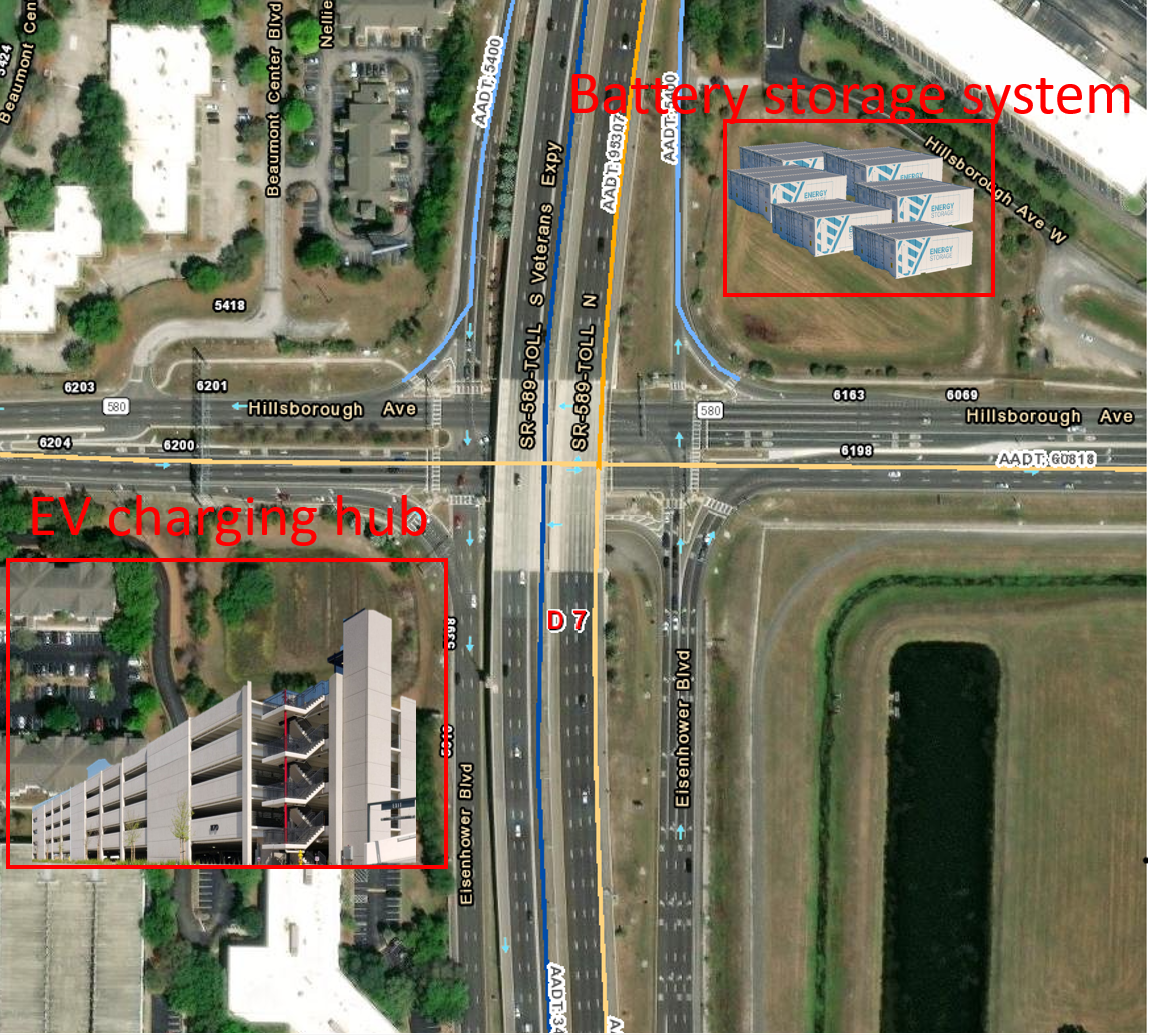}
    \caption{Hub and BSS location near a major street intersection in Tampa, Florida, USA, as considered in the numerical case study}
    \label{fig: intersection}
\end{figure}

We build a representative case study to demonstrate the effectiveness of our proposed model. The hub and the BSS are assumed to be connected to the same bus of the power network and hence are subjected to the same electricity prices. The BSS specification considered is similar to one of the battery storage systems operated by an electric power provider in the Tampa Bay area, Florida, USA. It is comprised of six compartments, each with a maximum storage capacity of 4000 kW. Each compartment maintains a minimum power of 500 kW and has a maximum charge/discharge rate of 3000 kW/hr. The charging hub is considered to be located near a busy roadway intersection in the Tampa Bay area (see figure \ref{fig: intersection}) and is assumed to house 150 fast charging stations capable of charging the EVs at the maximum rate of 100kW/hr. It is assumed that, to limit arbitrage by the hub, maximum DA commitment is capped at two times the hourly expected EV charging demand. 

\begin{figure}[ht]
     \centering
     
     \begin{subfigure}[b]{0.32\textwidth}
         \centering
         \includegraphics[width=\textwidth]{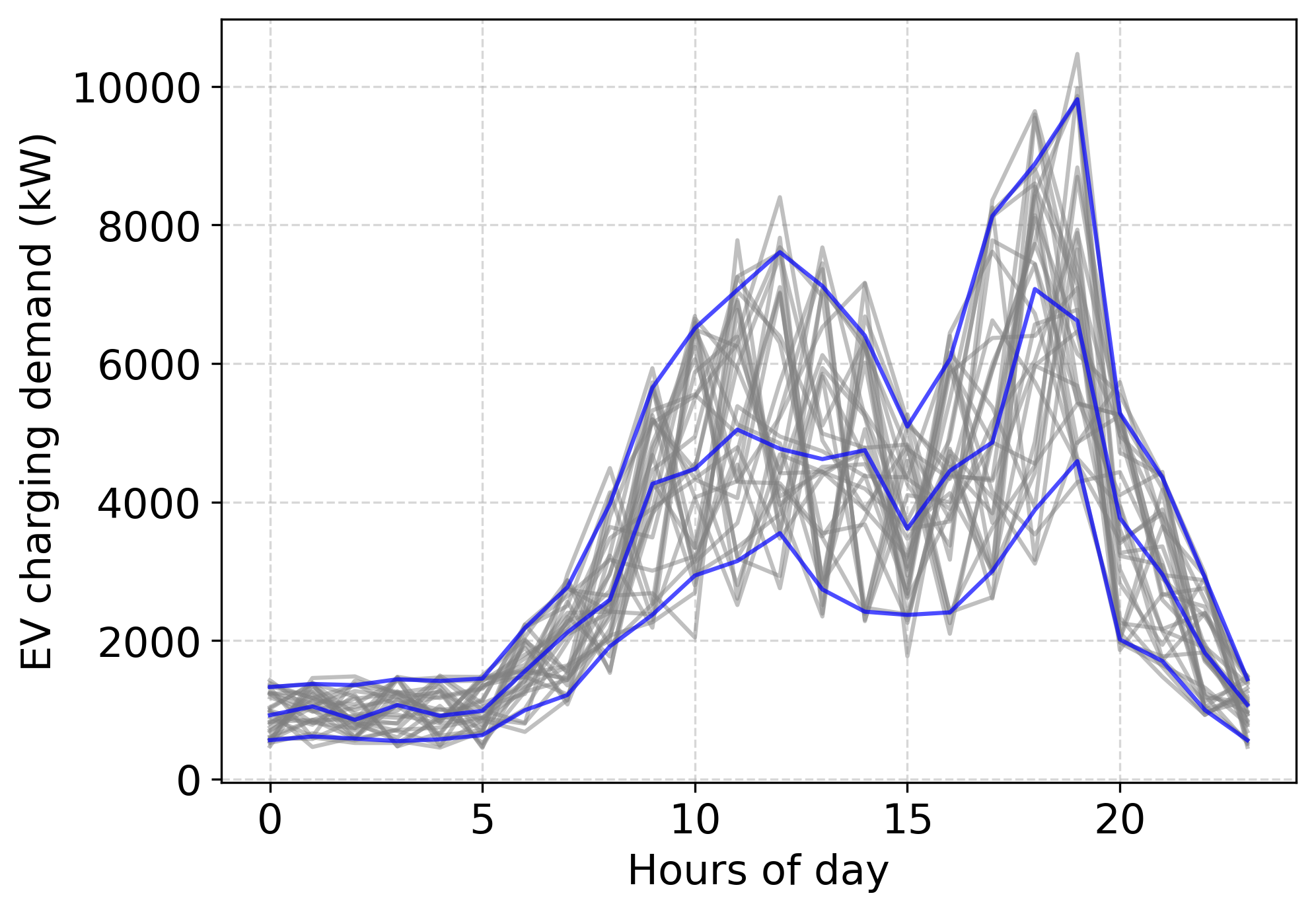}
         \caption{}
         \label{fig: ev demand}
     \end{subfigure}
     \hfill
     \begin{subfigure}[b]{0.32\textwidth}
         \centering
         \includegraphics[width=\textwidth]{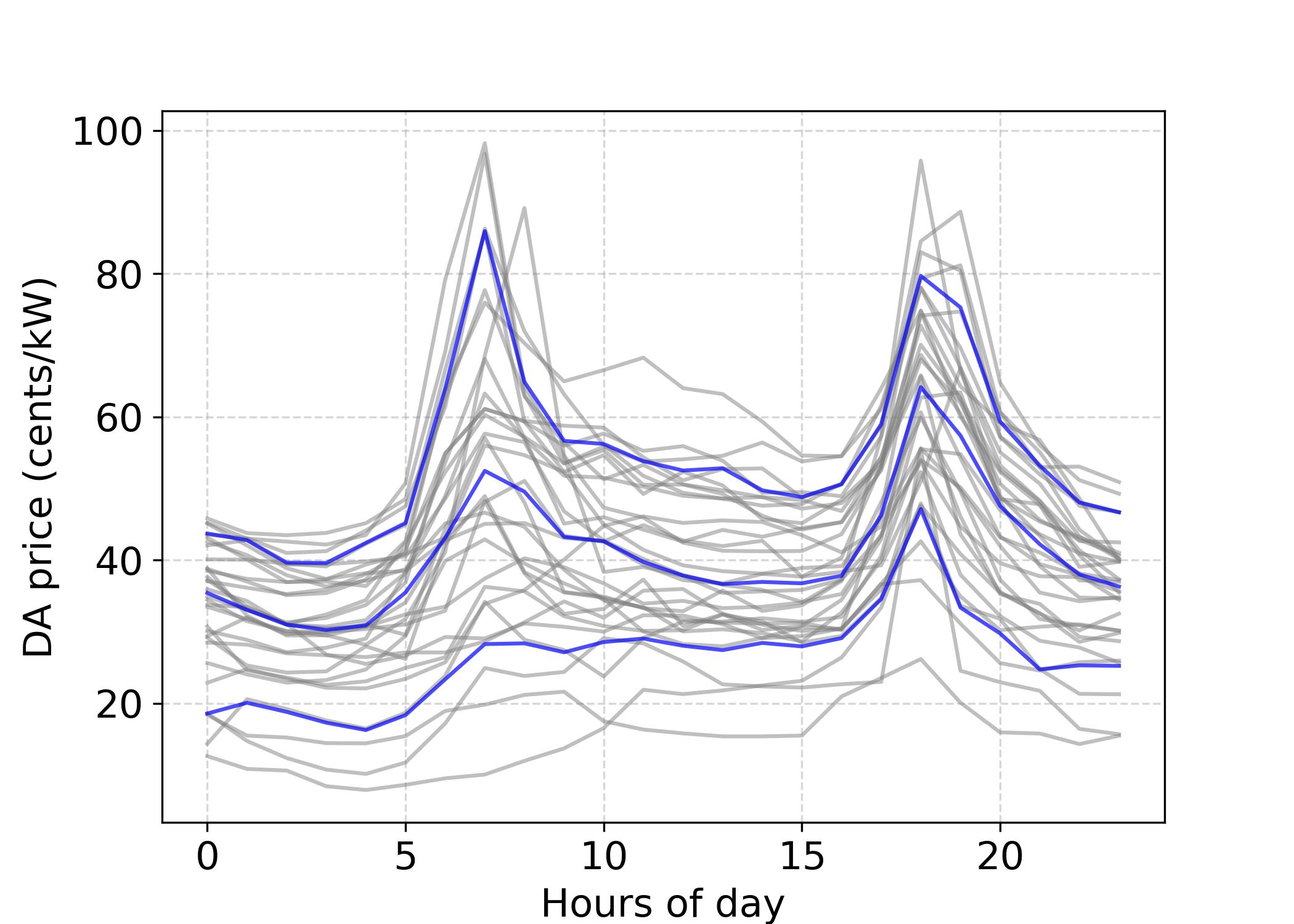}
         \caption{}
         \label{fig: da price}
     \end{subfigure}
     \hfill
     \begin{subfigure}[b]{0.32\textwidth}
         \centering
         \includegraphics[width=\textwidth]{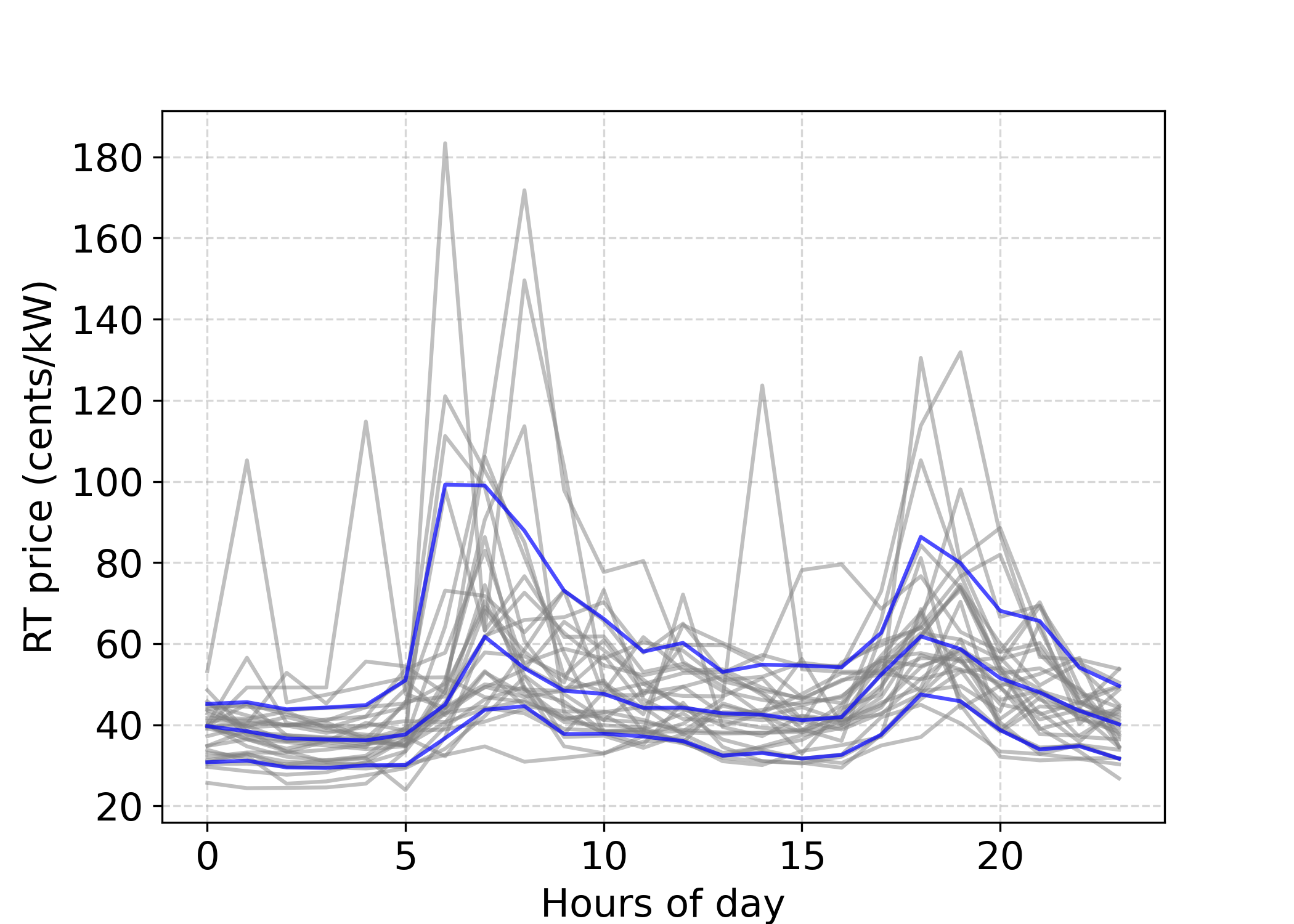}
         \caption{}
         \label{fig: rt price}
     \end{subfigure}
     
        \caption{(a) Total EV charging demands, (b) Day-ahead prices, and (c) Real-time prices}
        \label{fig: EV figures}
\end{figure}

The number of EVs seeking to charge in the hub at any hour is determined as follows. The hourly average traffic flow data through the intersection ($N_t$), as depicted in Figure \ref{fig: intersection}, is obtained from the Florida Department of Transportation \cite{traffic_flow}. We adopted traffic data for the month of September 2023. It is assumed that the $\alpha \%$ of the traffic are EVs and $\beta \%$ of those EVs use public charging facilities. Then the average number of the EVs passing through the intersection in each hour $t$ that might seek to charge at the hub is $\alpha \beta N_t$. These average values are used as the rate parameters for Poisson distributions to generate the hourly numbers of EVs that are potential candidates to charge in the hub $(\hat{N_t})$. The actual hourly number of EVs that charge in the hub $(n_t)$ is found by using a binomial distribution with parameters $( \hat{N_t}, p_t)$, where $p_t$ is the probability of an EV with potential to charge actually receiving charge. The values of $p_t$ are adopted from the EV charging behavior study by Idaho National Lab \cite{deng2018demand}. We consider $\alpha$ = 0.25 \cite{McKinsey} and  $\beta$ = 0.42 \cite{narayan2022dynamic}. EVs arriving at the hub are assumed to have three different battery sizes, 50 kW, 75 kW, and 100 kW with probabilities of 0.3, 0.4, and 0.3, respectively. As per the study conducted by Idaho National Laboratory in the U.S. \cite{Idaho}, the amount of charge that an EV seeks to receive varies between 5\% to 95\% of its battery size. The hourly values of the total demand for charge in the hub for all 30 days are shown in Figure \ref{fig: ev demand}. The blue lines in the figure represent the $10^{th}$, $50^{th}$, and $90^{th}$ percentiles. Since the Tampa Bay area electricity supply is regulated and does not have DA and RT markets, the hourly DA and RT prices are adopted from the electricity market (ERCOT) in the State of Texas, U.S. \cite{ERCOT}. The DA and RT price profiles are given in Figures  \ref{fig: da price} and \ref{fig: rt price}, respectively, with the $10^{th}$, $50^{th}$, and $90^{th}$ percentiles marked in blue.  

\begin{figure}[ht]
     \centering
     \begin{subfigure}[b]{0.32\textwidth}
         \centering
         \includegraphics[width=\textwidth]{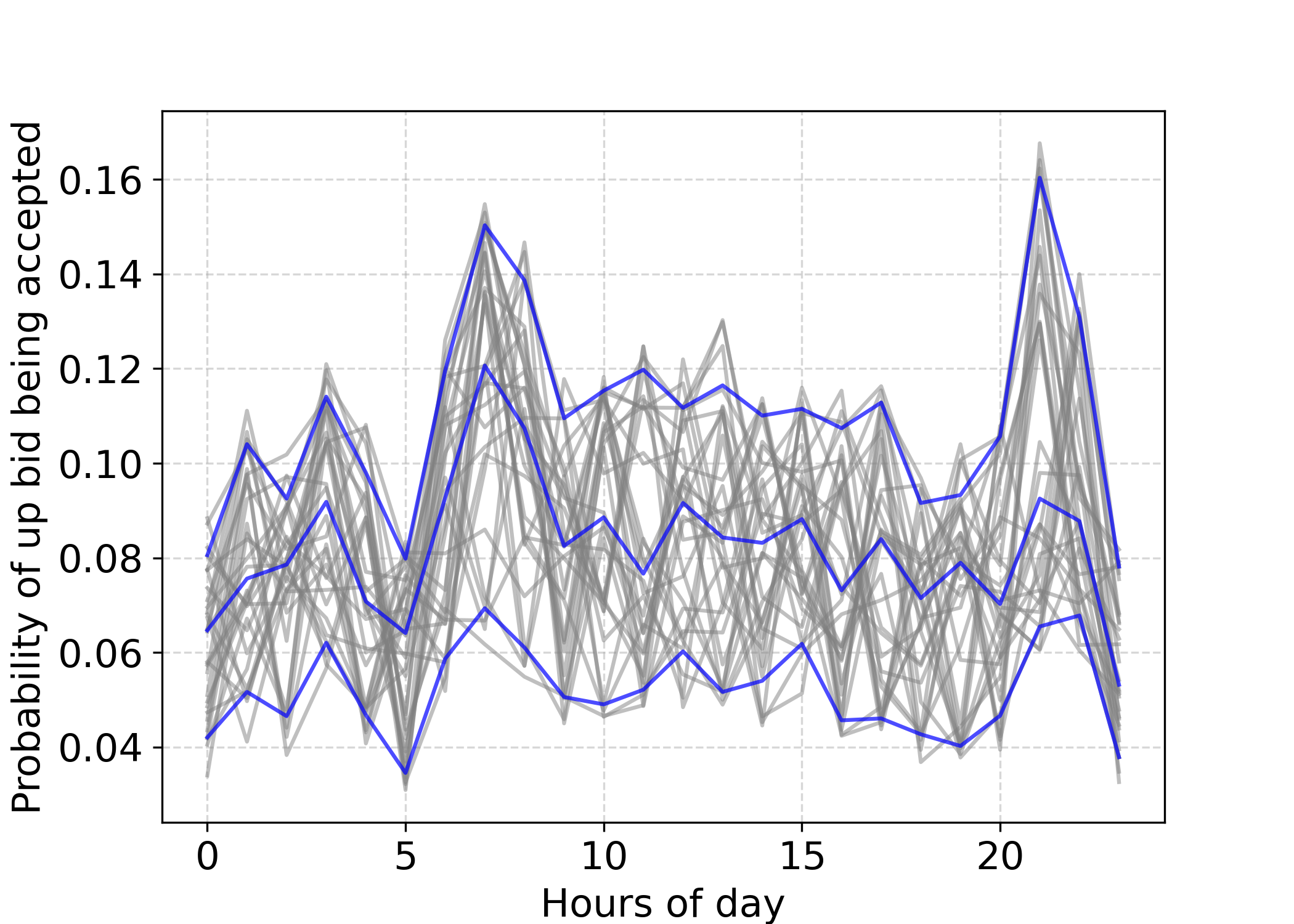}
         \caption{ Up bid acceptance probability}
         \label{fig: p_a_up}
     \end{subfigure}
     \hfill
     \begin{subfigure}[b]{0.32\textwidth}
         \centering
         \includegraphics[width=\textwidth]{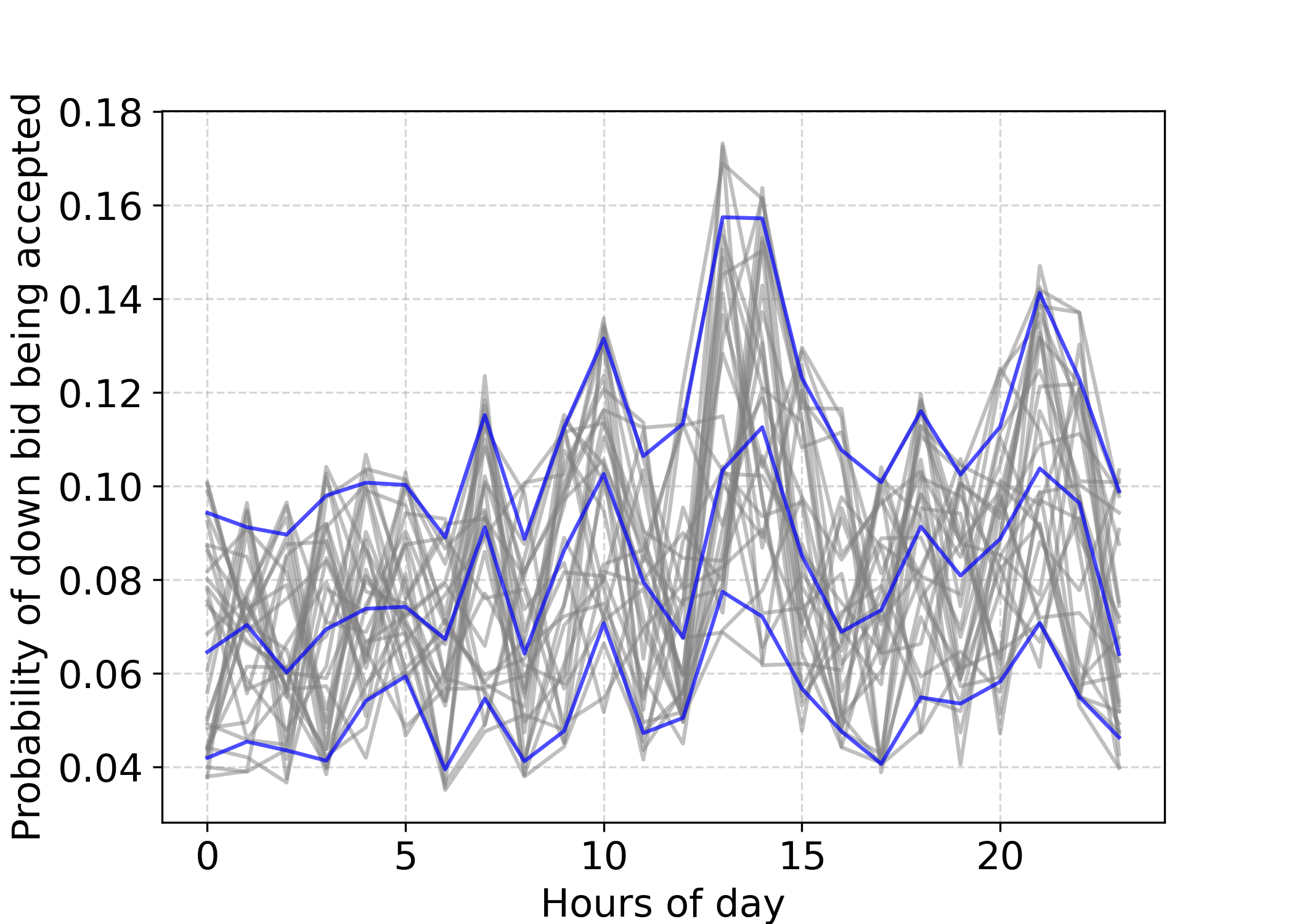}
         \caption{Down bid acceptance probability}
         \label{fig: p_a_dn}
     \end{subfigure}
     \hfill
     \begin{subfigure}[b]{0.32\textwidth}
         \centering
         \includegraphics[width=\textwidth]{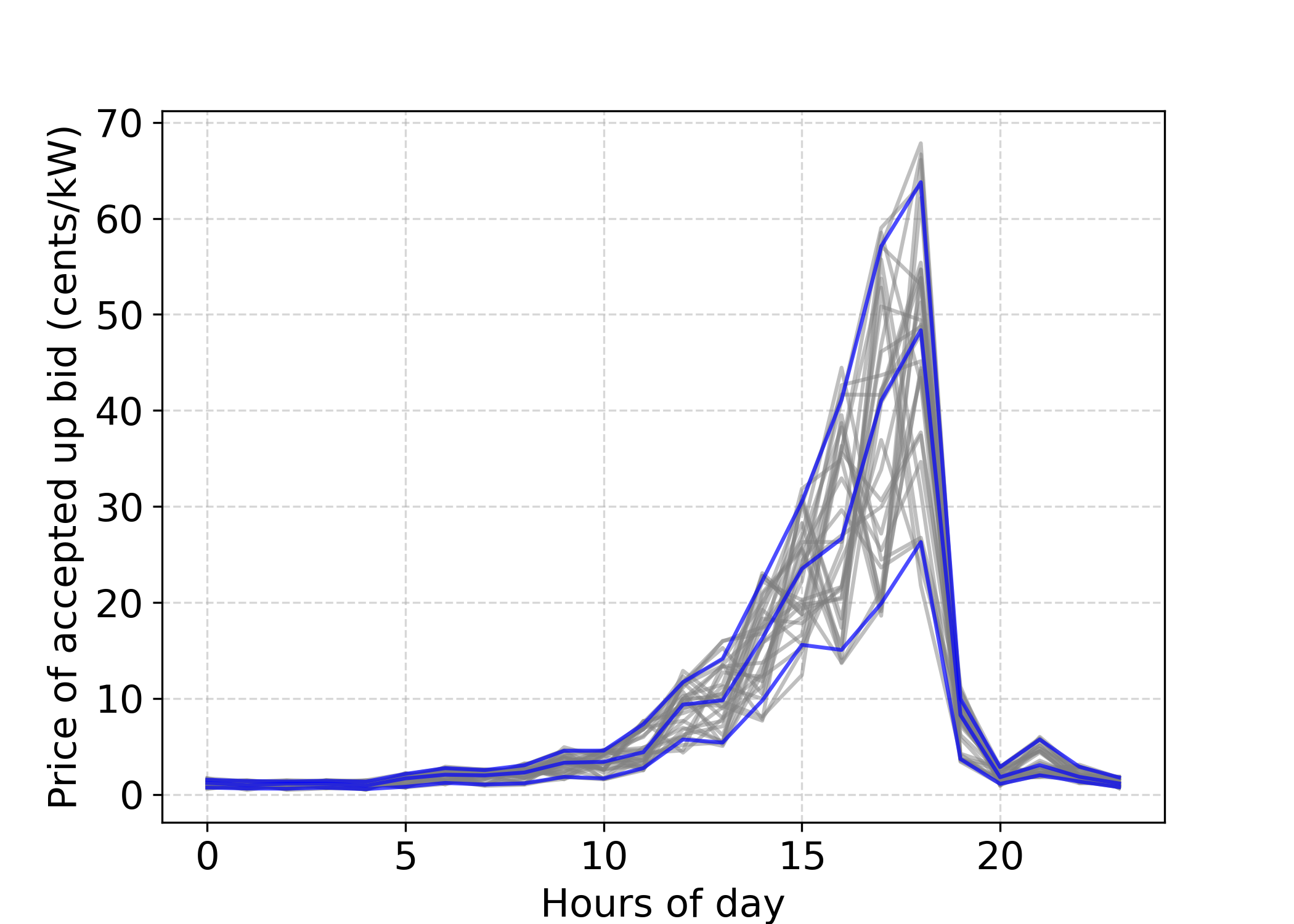}
         \caption{Accepted price of up bids}
         \label{fig: c_up}
     \end{subfigure}
     \hfill
     \begin{subfigure}[b]{0.32\textwidth}
         \centering
         \includegraphics[width=\textwidth]{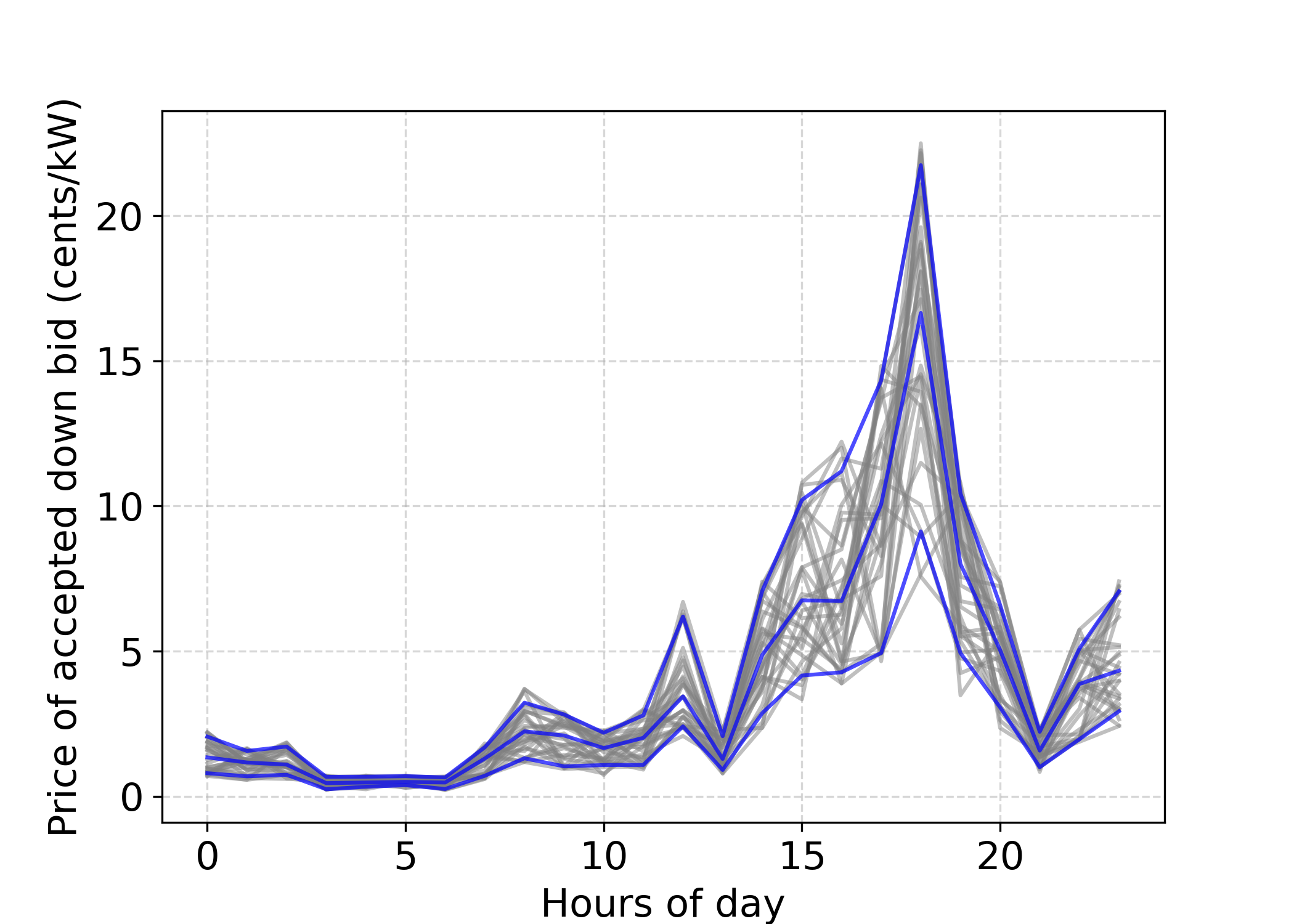}
         \caption{Accepted price of down bids}
         \label{fig: c_dn}
     \end{subfigure}
     \hfill
     \begin{subfigure}[b]{0.32\textwidth}
         \centering
         \includegraphics[width=\textwidth]{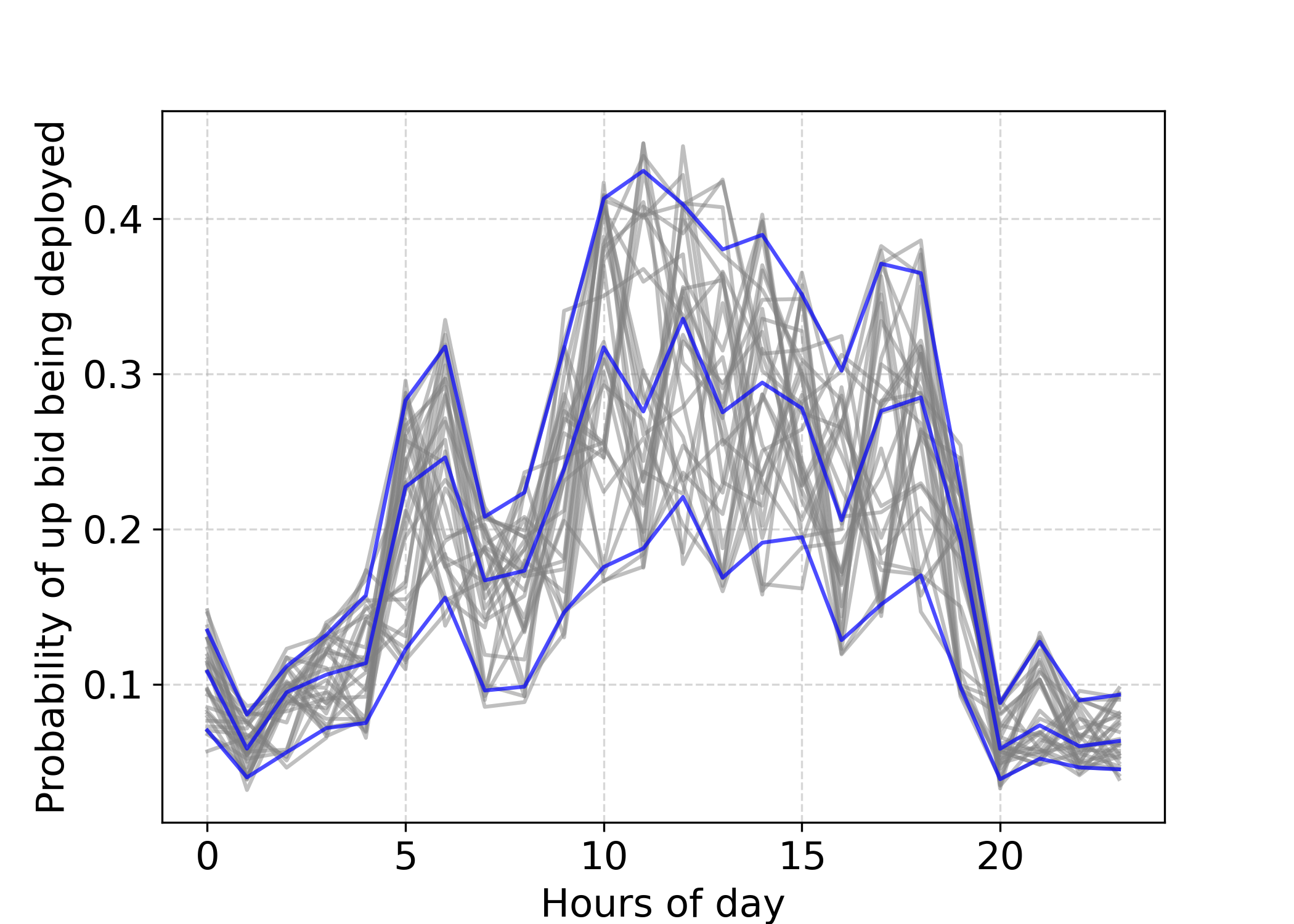}
         \caption{Deployment probability of up bids}
         \label{fig: p_d_up}
    \end{subfigure}
    \hfill
     \begin{subfigure}[b]{0.32\textwidth}
         \centering
         \includegraphics[width=\textwidth]{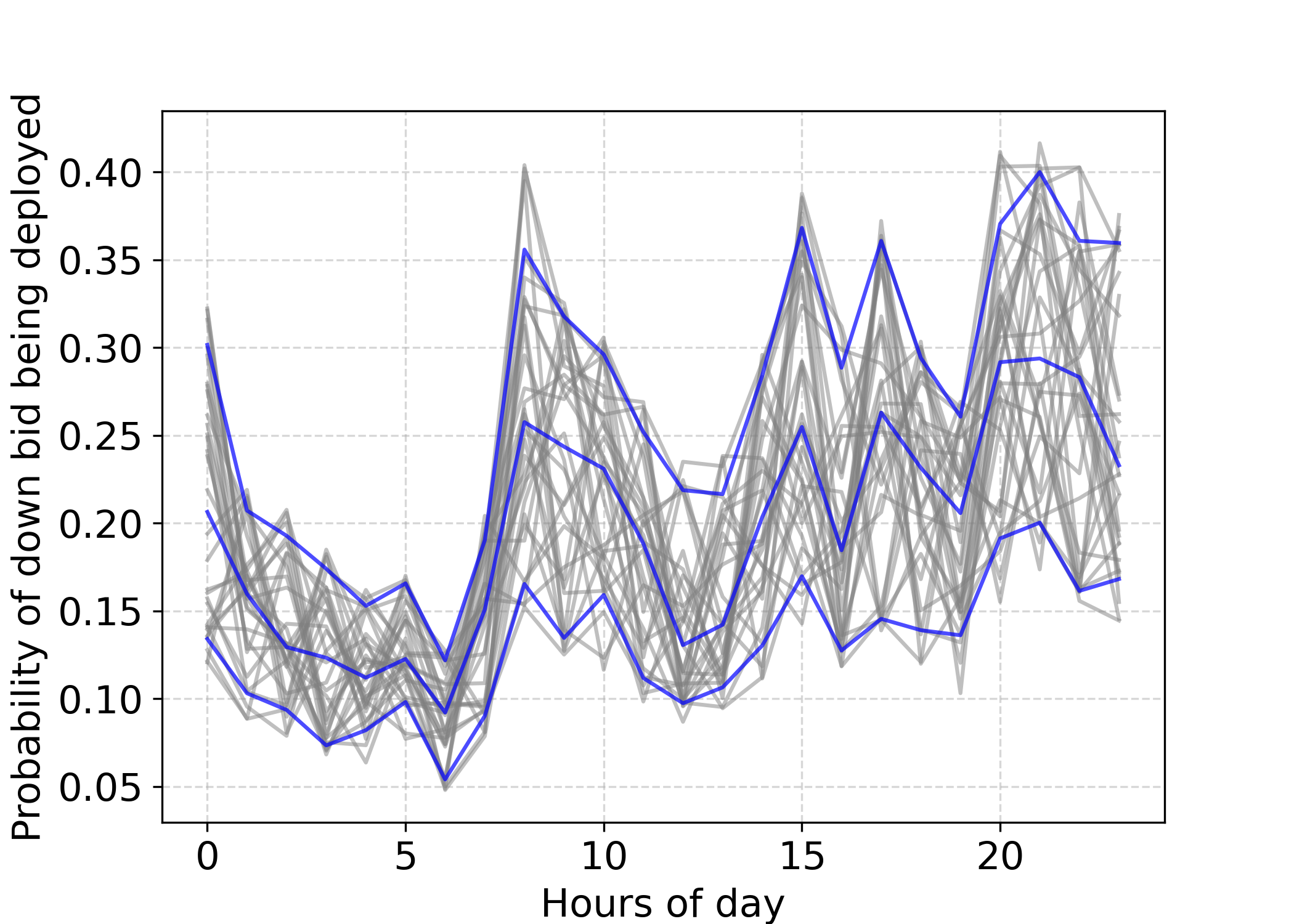}
         \caption{Deployment probability of down bids}
         \label{fig: p_d_dn}
     \end{subfigure}
        \caption{Results of reserve market clearing simulation using ERCOT data}
        \label{fig: simulation_of_ancillary_market}
\end{figure}

The battery storage system operator estimates the probability of acceptance of the reserve up and down bids, as well as the probability of deployment of the accepted bids via simulation of the market clearing mechanism. The simulation uses published historical data from the ERCOT market for the price/quantity bids for all market participants, total accepted quantities in the DA reserve market, and the total deployed up and down quantities. The simulation yields the market clearing prices and the accepted bids. The number of accepted bids divided by the total number of bids yields the probability of acceptance of a reserve bid. Similarly, the total quantity of reserve power deployed divided by the total quantity of accepted bids yields the probability of deployment. The simulation results are summarized in Figure \ref{fig: simulation_of_ancillary_market}.
Figures \ref{fig: p_a_up} and \ref{fig: p_a_dn} depict the probability of acceptance of reserve up and reserve down bids, respectively. The price of accepted up and down bids are shown in figure \ref{fig: c_up} and \ref{fig: c_dn}, respectively. The deployment probability for the accepted reserve up and down bids are shown in figures \ref{fig: p_d_up} and \ref{fig: p_d_dn}, respectively. 

The mathematical models are solved using Gurobi 9.5.2 in Python 3.8 using a computer with an Intel i9-11900H@2.50GHz processor supported by 32 GB RAM. All the models are solved to an optimality gap of 0.05\% or less. In what follows, we describe the scenarios for which the optimization models (P1, P2, and P3, described earlier) are solved and then present the corresponding results. 

\subsection{Results}

We first solve our models P1, P2, and P3 with all the input parameters set at their median values as shown in Figures \ref{fig: EV figures} and \ref{fig: simulation_of_ancillary_market}. We solve P3 using a Nash bargaining approach (as presented in \eqref{eq: Nash SOCP}). We also solve P3 using a total cost minimization (TCM) approach \eqref{eq: TCM} to demonstrate the benefit of the NBS in terms of objective function value fairness for both the hub and the BSS in the joint operation mode. In the TCM approach, we minimize the combined cost of the hub and the BSS (negative of its profit) given as  
\begin{align}\label{eq: TCM}
    \begin{split}
        \min_{\bm{z}} f_{p3}^a - f_{p3}^b \\
        \bm{z} \in \bm{Z_f},
    \end{split}
\end{align}
which is a MILP and can be solved using any commercial solver. The results are shown in Table \ref{table: tcm_vs_nbs}. When the hub and the BSS act independently (as in P1 and P2, respectively), both the hub cost and BSS profit are inferior to those obtained from the joint operation (TCM and NBS). In the TCM solution, the hub being a larger entity influences the solution and receives the most benefit from cooperation with 11.02\% cost reduction while the BSS increases its profit only by 0.43\% compared to the solutions from independent operations. Whereas, the NBS offers a fair distribution of the benefits of cooperation with a 7.32\% cost reduction for the hub and a 5.23\% profit increase for the BSS. 

\begin{table}[ht]
\begin{tabular}{c|c|c|c|}
\cline{2-4}
                                      & Independent Operation & Joint Operation per TCM (P3)                & Joint Operation per NBS (P3)                \\ \hline
\multicolumn{1}{|l|}{Hub cost (\$)}   & 30,262.87 (from P1)  & 26,929.22 (-11.02\%) & 28,048.70 (-7.32\%) \\ \hline
\multicolumn{1}{|l|}{BSS profit (\$)} & 7,073.65 (from P2)  & 7,103.95 (0.43\%)    & 7,443.91 (5.23 \%)  \\ \hline
\end{tabular}
\caption{Hub costs and BSS profits}
\label{table: tcm_vs_nbs}
\end{table}

\begin{figure}[hbt!]
    \centering
\includegraphics[width=1\linewidth]{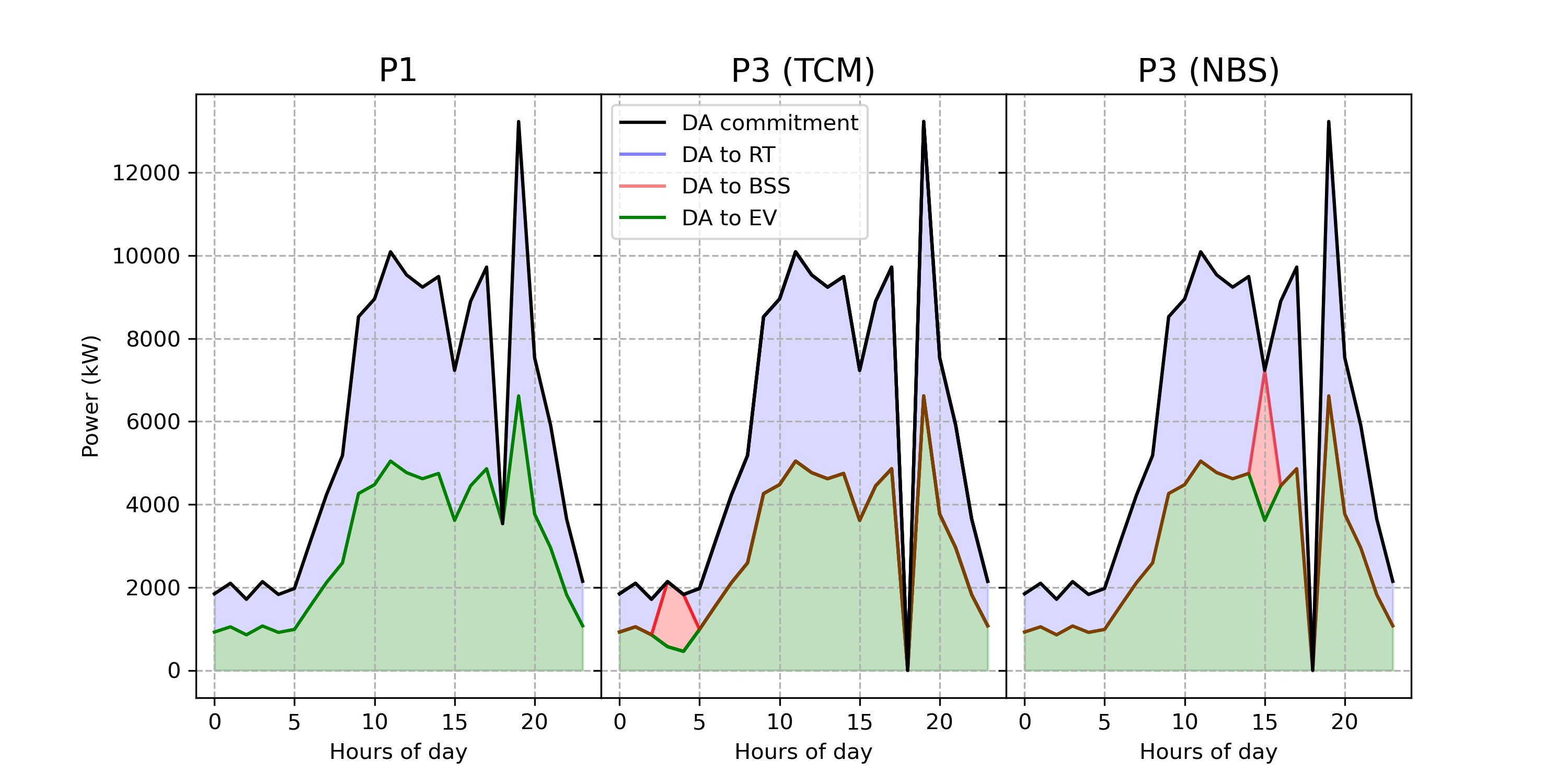}
    \caption{DA commitment and utilization at the median values of model parameters}
    \label{fig: da_com}
\end{figure}

\begin{figure}[hbt!]
    \centering
\includegraphics[width=1\linewidth]{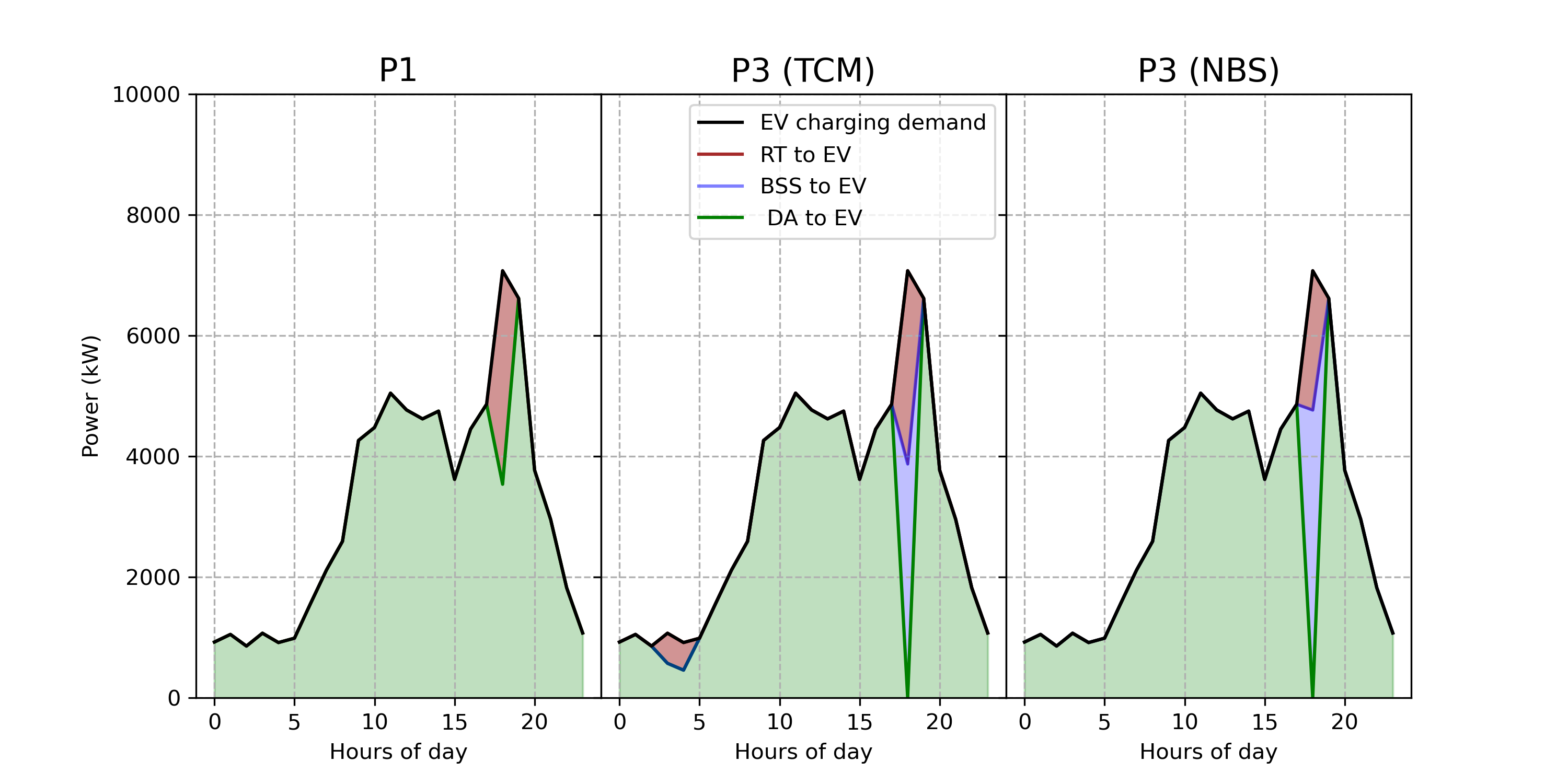}
    \caption{EV charging demand and sources for meeting the demand at median values of model parameters}
    \label{fig: EV charging demand}
\end{figure}

\begin{figure}[ht]
    \centering
\includegraphics[width=1\linewidth]{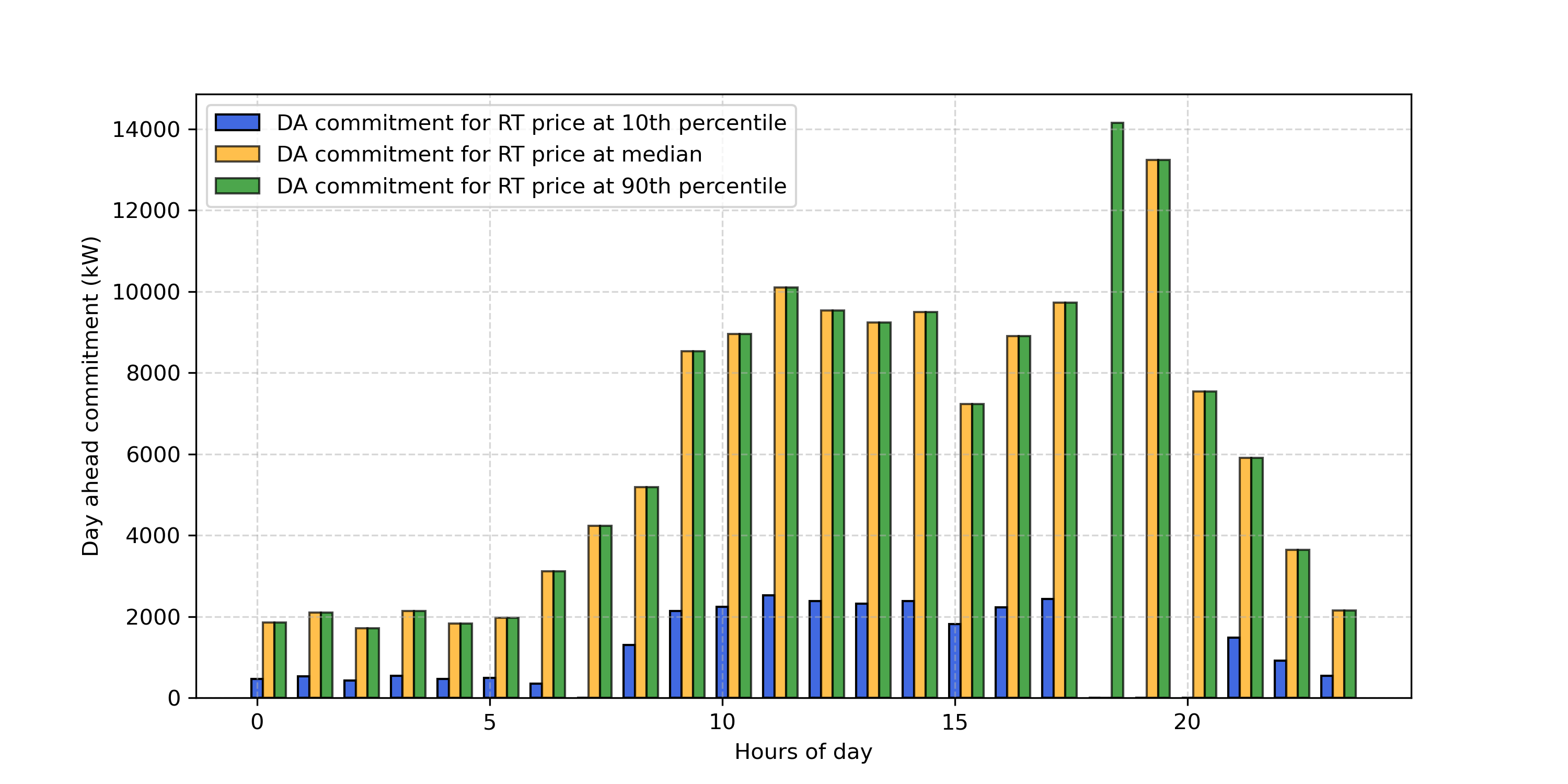}
    \caption{DA commitment with different RT prices with DA price at the median}
    \label{fig: sensitivity_rt_price_da_com}
\end{figure}

Since DA commitment is a primary source to procure power by the hub to control cost in meeting the charging demand, we examined the DA commitment behavior and DA power usage in independent as well as joint operation modes. Figure \ref{fig: da_com} shows the hourly DA commitments and DA usage when all the model parameters (prices, charging demand, and reserve market probabilities) are at their median values. The outer boundary line of each graph is obtained by joining the hourly DA commitment values. The shaded areas under the line represent the amounts of DA power used for charging EVs, storing in the BSS, and selling back to the RT market. We note that since the median RT prices are generally higher than DA prices, the hub's choices for DA commitment in all operation modes are similar, except for hour 18. It can be observed that the hub generally commits to more than the EV charging requirement and sells the excess mostly to the RT market and only sparingly stores DA power in the BSS. Figure \ref{fig: EV charging demand} also shows that when all parameters are at their median values, most of the EV charging power is sourced from the DA market except for a few hours when charging is supplemented by power from RT and BSS. Figure \ref{fig: sensitivity_rt_price_da_com} depicts the DA commitment pattern for various RT prices when DA price and other parameters are their median values. As expected, when the RT price is lower than DA (RT at the 10th percentile and DA at the median), the hub chooses to minimize its DA commitment. In both of the other cases, RT prices are higher than DA and hence the DA commitments are similar.

\begin{figure}[hbt!]
    \centering
\includegraphics[width=1\linewidth]{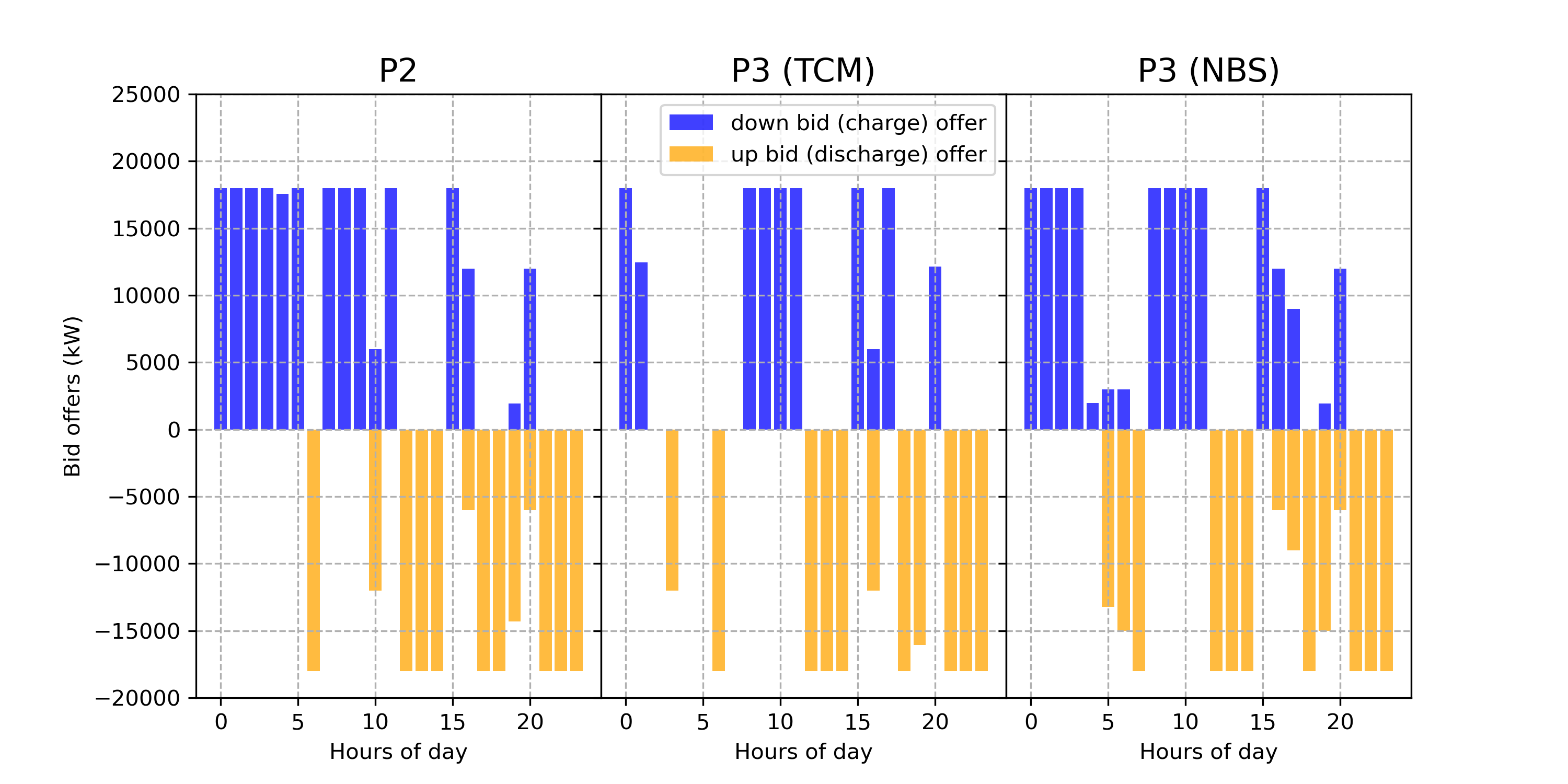}
    \caption{Reserve bid patterns by BSS} 
    \label{fig: reserve bid}
\end{figure}

\begin{figure}[hbt!]
    \centering
\includegraphics[width=1\linewidth]{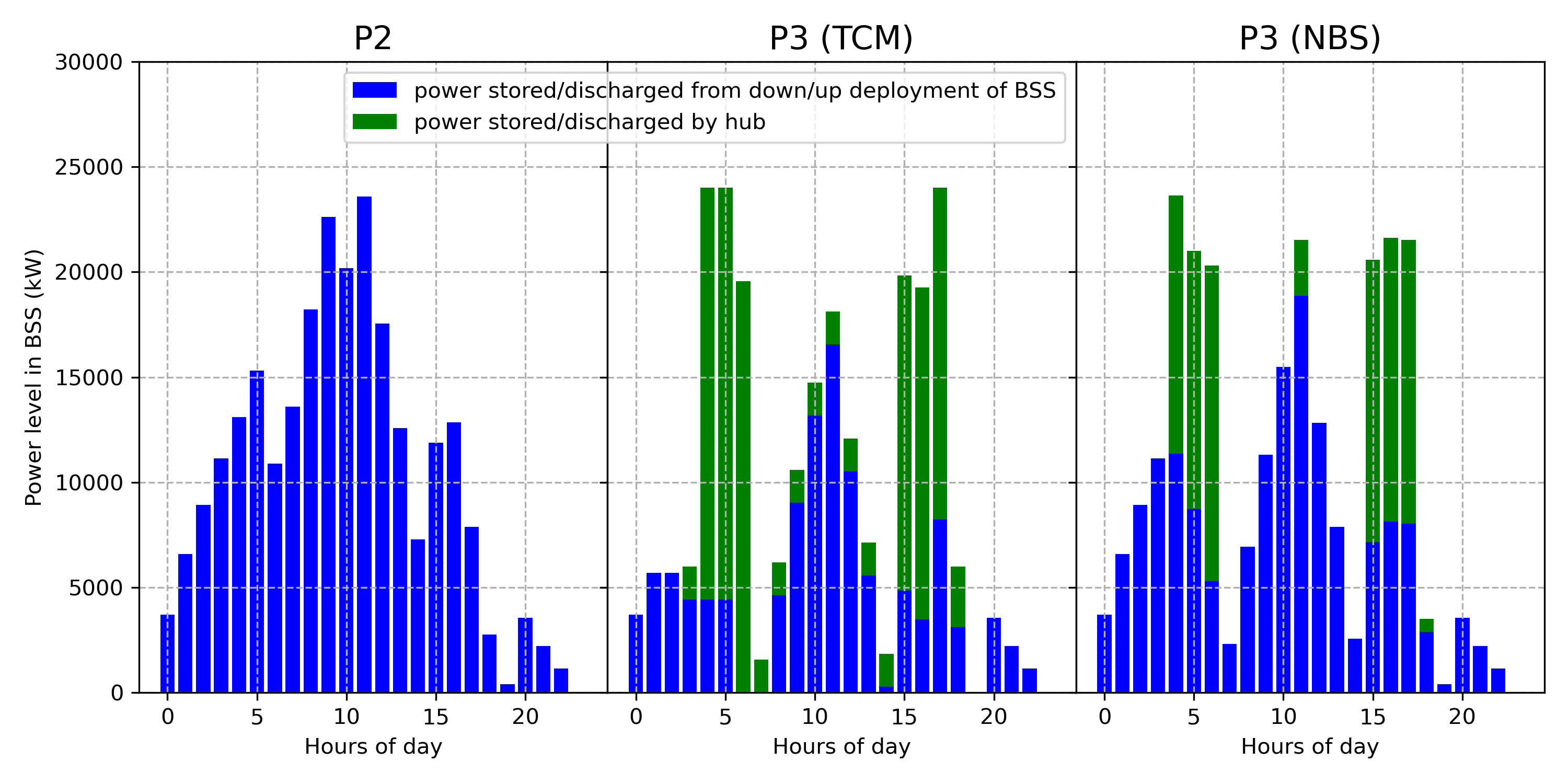}
    \caption{BSS power}
    \label{fig: bss_power}
\end{figure}

Since the BSS aims to maximize its profit by cooperating with the hub while also participating in the reserve market, we examined BSS' reserve market bidding behavior in independent and joint operation modes (see Figure \ref{fig: reserve bid}). It can be seen that when BSS cooperates with the hub (P3), it offers less bid to the reserve market compared to its independent operation (P2), which allows for the use of BSS storage capacity by the hub. It can also be noted that in the TCM approach, the hub, being the larger entity, skews the solution in its favor by using a larger share of the storage capacity, thus reducing the bid offers by the BSS. In the more equitable approach (NBS), the BSS is allowed to use more of its capacity, when appropriate, to increase its profit, and hence we see more bid offers compared to TCM. We also note that BSS makes more down bids in the early hours of the day to store power, and submits more up bids in later hours to discharge when prices are higher. Figure \ref{fig: bss_power} depicts the BSS capacity usage under independent and joint operation modes. Similar to that noted from Figure \ref{fig: reserve bid}, the NBS solution in Figure \ref{fig: bss_power} allows for more fair usage of storage capacity by the BSS as opposed to being dominated by the hub, as in the TCM solution.

\newpage
\subsection{Sensitivity Analysis}
The hub and the BSS while operating jointly guided by the NBS must consider a multitude of parameters (including electricity and reserve market prices, EV charging demand, and down/up acceptance and deployment probabilities) in making respective cost-minimizing and profit-maximizing decisions. Even if we consider only three values for each of the parameters (low (10th percentile), median, and high (90th percentile)), there are $3^9$ possible combinations to examine. Hence, we group the parameters into two subsets, each having a more direct influence on a decision-maker. 

\begin{table}[htp]
\begin{tabular}{
>{\columncolor[HTML]{C0C0C0}}c lccccccccc}
\multicolumn{11}{c}{\cellcolor[HTML]{C0C0C0}\textbf{EV charging demand}}                                                                     \\ \hline
\cellcolor[HTML]{C0C0C0}                                    & \multicolumn{1}{l|}{}                   & \multicolumn{3}{c|}{Low}                                                               & \multicolumn{3}{c|}{Median}                                                            & \multicolumn{3}{c}{High}                                         \\ \cline{3-11} 
\cellcolor[HTML]{C0C0C0}                                    & \multicolumn{1}{l|}{}                   & \multicolumn{3}{c|}{\cellcolor[HTML]{C0C0C0}\textbf{RT price}}                         & \multicolumn{3}{c|}{\cellcolor[HTML]{C0C0C0}\textbf{RT price}}                         & \multicolumn{3}{c}{\cellcolor[HTML]{C0C0C0}\textbf{RT price}}    \\ \cline{3-11} 
\cellcolor[HTML]{C0C0C0}                                    & \multicolumn{1}{l|}{\multirow{-3}{*}{}} & \multicolumn{1}{c|}{Low}   & \multicolumn{1}{c|}{Median} & \multicolumn{1}{c|}{High}   & \multicolumn{1}{c|}{Low}   & \multicolumn{1}{c|}{Median} & \multicolumn{1}{c|}{High}   & \multicolumn{1}{c|}{Low}   & \multicolumn{1}{c|}{Median} & High  \\ \cline{2-11} 
\cellcolor[HTML]{C0C0C0}                                    & \multicolumn{1}{l|}{Low}                & \multicolumn{1}{c|}{4.17}  & \multicolumn{1}{c|}{42.54}  & \multicolumn{1}{c|}{170.87} & \multicolumn{1}{c|}{2.50}  & \multicolumn{1}{c|}{25.32}  & \multicolumn{1}{c|}{110.45} & \multicolumn{1}{c|}{1.75}  & \multicolumn{1}{c|}{17.93}  & 76.90 \\ \cline{2-11} 
\cellcolor[HTML]{C0C0C0}                                    & \multicolumn{1}{l|}{Median}             & \multicolumn{1}{c|}{5.16}  & \multicolumn{1}{c|}{11.92}  & \multicolumn{1}{c|}{52.96}  & \multicolumn{1}{c|}{4.35}  & \multicolumn{1}{c|}{7.32}   & \multicolumn{1}{c|}{31.97}  & \multicolumn{1}{c|}{3.76}  & \multicolumn{1}{c|}{5.14}   & 22.33 \\ \cline{2-11} 
\multirow{-6}{*}{\cellcolor[HTML]{C0C0C0}\textbf{DA price}} & \multicolumn{1}{l|}{High}               & \multicolumn{1}{c|}{12.02} & \multicolumn{1}{c|}{11.33}  & \multicolumn{1}{c|}{23.04}  & \multicolumn{1}{c|}{11.67} & \multicolumn{1}{c|}{7.58}   & \multicolumn{1}{c|}{14.01}  & \multicolumn{1}{c|}{10.63} & \multicolumn{1}{c|}{5.76}   & 9.78 
\end{tabular}
\caption{Sensitivity of DA and RT prices and EV charging demand on hub's cost reduction (in percentage) from NBS compared to its individual solution (P1)}
\label{table: sensitivity of hub}
\end{table}

We examine the sensitivity of DA and RT prices and EV charging demand on the hub's cost by considering three possible values (low, median, and high) of the parameters, which resulted in $3^3$ different combinations. For each of these combinations, all other model parameters that are concerned with the reserve market and hence primarily affecting the BSS' profit are kept at their median values. The sensitivity is measured by the percentage reduction in the hub's NBS cost compared to its cost when operating independently (P1). This is calculated as \{(NBS cost - P1 cost)/(P1 cost)\}$\times$ 100 and are presented in Table \ref{table: sensitivity of hub}. Since the NBS differs from the P1 solution due to its ability to arbitrage using BSS' storage capacity, all the numbers in the table reflect the incremental cost reduction benefits by the increase in revenue from arbitrage. Clearly, for some DA/RT prices and EV charging demand combinations, the arbitrage potentials for the hub are either very high or very low. The results show that the very high potentials are generally for the combinations when the DA prices are low and RT prices are high. The arbitrage potential is low when both the DA and RT prices are low. It may also be noted that the hub's arbitrage potential decreases, irrespective of DA and RT price combinations, with an increase in EV charging demand. 

\begin{figure}[hbt!]
    \centering
\includegraphics[width=1\linewidth]{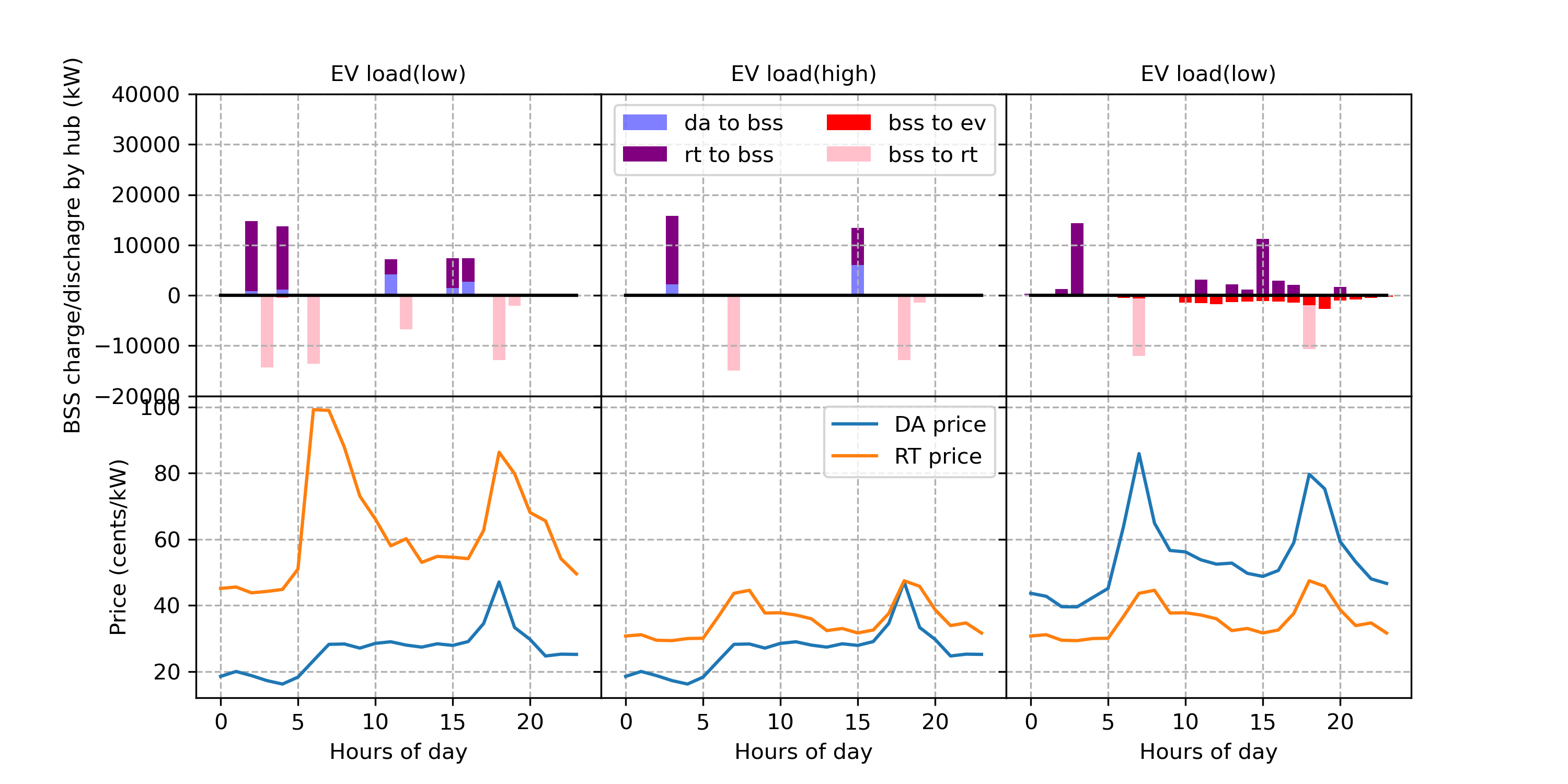}
    \caption{BSS usage by the hub under different DA prices, RT prices, and EV charging demand}
    \label{fig: extreme cases}
\end{figure}

Among the combinations studied in Table \ref{table: sensitivity of hub}, the highest cost reduction potential (170.87 \%) occurs when the DA price is low, the RT price is high, and the EV charging demand is low. To further examine this case, we looked at the BSS usage by the hub across all hours of the day. This is depicted in the leftmost part of Figure \ref{fig: extreme cases}, where the line graphs show RT prices to be significantly higher than the DA prices, and the bar graphs show multiple charge/discharge activities by the hub. We note that the significant price differential between DA and RT prices encourages the hub operator to increase its level of arbitrage via high usage of BSS storage capacity and thus reduce its cost. We also examine two other cases: low DA and RT prices and high EV charging demand (cost reduction of 1.75\%); and high DA prices, low RT prices, and low EV charging demand (cost reduction of 12.02\%). The BSS usage by the hub for these two cases is depicted in the middle and the rightmost segment of Figure \ref{fig: extreme cases}, respectively. For the case with a price reduction of 1.75\%, the low differential between DA and RT prices prompts a limited hub engagement with the BSS (i.e. lower level of charge/discharge activity) yielding a lower potential for cost reduction.  Whereas in the case of a 12.02 \% cost reduction, we note that the RT prices are significantly lower than the DA prices and also do not vary widely across the hours. This yields a moderate amount of arbitrage by the hub using the BSS capacity. 

\begin{table}[ht]
\begin{tabular}{|c|c|c|c|c|c|}
\rowcolor[HTML]{C0C0C0} 
Parameters                                                                                                              & F-statistic & F-critical & P-value  & Decision                \\ \hline
Up bid market clearing price $(\lambda^{up})$                                                                                          &  11.92       & 4.279      & 0.002     &       Significant             \\
\hline
Down bid market clearing price $(\lambda^{dn})$                                                                                       &  1.82       & 4.279      & 0.18      &    Not Significant               \\
\hline
Probability of acceptance of up bid  $(\pi^{a,up})$                                                                                   &  8.82        & 4.279      & 0.006      &       Significant             \\ 
\hline
Probability of acceptance of down bid  $(\pi^{a,dn})$                                                                                    &  0.96        & 4.279      & 0.33      &      Not Significant              \\ 
\hline
Probability of deployment of up bid  $(\pi^{d,up})$                                                                                   &  83.79       & 4.279      & 3.94 $\times$ $10^{-9}$ & Significant \\ \hline
Probability of deployment of down bid      $(\pi^{d,dn})$                                                                             &  32.10       & 4.279      & 9.07 $\times$ $10^{-6}$ & Significant \\ \hline
\begin{tabular}[c]{@{}c@{}}Up bid market clearing price $(\lambda^{up})$ $\times$ \\ probability of deployment  of up bid $(\pi^{d,up})$ \end{tabular}            & 6.210       & 4.279      & 0.020    & Significant                      \\ \hline
\begin{tabular}[c]{@{}c@{}}Probability of deployment of up bid $(\pi^{d,up})$ $\times$ \\ probability of deployment of down bid $(\pi^{d,dn})$ \end{tabular}     & 6.36        & 4.279      & 0.019    &    Significant  \\
\hline
\end{tabular}
\caption{Analysis of variance (ANOVA) to examine the sensitivity of reserve market parameters on BSS's profit increase (in percentage) from NBS compared to its individual solution (P2)}
\label{table: sensitivity of bss}
\end{table}

Hereafter, we study the sensitivity of six reserve market parameters on the BSS's ability to increase profit by allowing the hub to use its storage capacity. This profit increase potential (in percentage) is calculated as \{(NBS profit - P2 profit)/ P2 profit\} $\times$ 100 \%. The parameters studied are up and down bid market clearing prices, probabilities of acceptance of up and down bids, and probabilities of deployment of up and down bids. We used a $2^{6-1}$ fractional factorial design-based analysis of variance (ANOVA) study, see chapter 8 of \cite{montgomery2017design} for the details of $2^k$ factorial design. The 10th and 90th percentile values are considered as the two levels of the parameters and the highest interaction term is used as the fractional factorial design generator. In the thirty-two experiments for the $2^{6-1}$ fractional factorial design, the percentage profit varied between 1.52\% and 45\%. The ANOVA results are presented in Table \ref{table: sensitivity of bss}. Of the six reserve market parameters, the down-bid market clearing price and its probability of acceptance are found not significant with a type-I error level of 5\%. We also found two of the two-parameter interactions to be significant, although with a small margin. Of the four statistically significant parameters, the probabilities of deployment of up and down bids are the most significant ones with very high F-statistic values of 83.79 and 32.10, respectively. This is expected as the actual up-and-down deployments are the major contributor to BSS revenue from the reserve market. Examination of the BSS behavior from NBS reveals that for high values of up and down deployment probabilities, the BSS uses a higher proportion of its storage capacity for the reserve market, leaving a lower capacity for the hub.

\section{Concluding remarks}\label{sec: conclusions}

In this paper, we develop a bi-objective joint operation model between a cost-minimizing fast-charging EV hub and a profit-maximizing grid-connected stand-alone battery storage system. The BSS primarily uses its capacity to participate in the reserve market. It also cooperates with the hub by allowing the hub to use its storage capacity for arbitrage. The bi-objective problem is reframed as a Nash bargaining problem, which is then reformulated into a second-order cone program. 
We develop a sample case study problem adopting DA and RT electricity prices from the ERCOT market and generating simulated values for the reserve market prices and their probabilities as well as the EV charging demand. We demonstrate the benefits/fairness of our proposed Nash bargaining solution with the solution from a total cost minimization approach as well as the independent solutions without cooperation. A detailed study of parameter sensitivity reveals operational guidance for both the hub and the BSS under various parameter value combinations.

Our work demonstrates that the EV charging hubs can significantly benefit financially by utilizing the storage capacity of grid-connected stand-alone BSS. We also show that the joint operation under NBS yields added benefits for the BSS. The results from the numerical case study, for a hub with 150 DC fast-charging stations and a BSS with 24 MW storage capacity, show that under specific operating conditions, the hub and the BSS can financially gain up to 170\% and 45\%, respectively. These findings create new opportunities for the growth of stand-alone BSS in support of large-scale fast-charging hubs, ultimately increasing the rate of EV adoption.

Our study has the following limitations. Our model does not explicitly consider uncertainties in the energy market and the reserve market parameters. We investigated the performance of our model for various parameter value combination scenarios. A distributionally robust approach to the Nash bargaining problem could be a way to further accommodate these uncertainties \cite{liu2018distributionally} \cite{peng2021games}. In our problem, we have considered participation in the reserve up and down markets as the primary activities of the battery storage system.  Whereas in most electricity markets, BSS also participates in other reserve market activities such as responsive reserve, non-spinning responsive reserve, and contingency reserve. Consideration of these additional activities will add more variables and constraints without changing the fundamental nature of the model as we have developed. For simplicity, we have considered only the hourly variation in the RT market prices in formulating our model, which can be extended to accommodate more frequent changes in the RT market.

\bibliographystyle{unsrtnat}
\bibliography{references}  

\begin{thebibliography}{29}
\providecommand{\natexlab}[1]{#1}
\providecommand{\url}[1]{\texttt{#1}}
\expandafter\ifx\csname urlstyle\endcsname\relax
  \providecommand{\doi}[1]{doi: #1}\else
  \providecommand{\doi}{doi: \begingroup \urlstyle{rm}\Url}\fi

\bibitem[Conzade et~al.(2021)Conzade, Cornet, Hertzke, Hensley, Heuss, Möller, Schaufuss, Schenk, Tschiesner, and Laufenberg]{McKinsey}
Julian Conzade, Andreas Cornet, Patrick Hertzke, Russell Hensley, Ruth Heuss, Timo Möller, Patrick Schaufuss, Stephanie Schenk, Andreas Tschiesner, and Karsten~von Laufenberg.
\newblock Why the automotive future is electric.
\newblock \url{https://www.mckinsey.com/industries/automotive-and-assembly/our-insights/why-the-automotive-future-is-electric}, 2021.
\newblock (accessed: 26 July 2022).

\bibitem[Tesla()]{Tesla}
Tesla.
\newblock Supercharger.
\newblock \url{https://www.tesla.com/en_eu/supercharger}.
\newblock (accessed : 2 July 2022).

\bibitem[NREL(2023)]{NREL}
NREL.
\newblock Utility-scale battery storage.
\newblock \url{https://atb.nrel.gov/electricity/2023/utility-scale_battery_storage}, 2023.
\newblock (accessed: 01 December 2023).

\bibitem[Zheng et~al.(2020)Zheng, Yu, Shao, and Jian]{zheng2020day}
Yanchong Zheng, Hang Yu, Ziyun Shao, and Linni Jian.
\newblock Day-ahead bidding strategy for electric vehicle aggregator enabling multiple agent modes in uncertain electricity markets.
\newblock \emph{Applied Energy}, 280:\penalty0 115977, 2020.

\bibitem[Paudel et~al.(2023)Paudel, Bustos, and Das]{paudel2023distributionally}
Diwas Paudel, Nicolas Bustos, and Tapas~K Das.
\newblock A distributionally robust approach for day-ahead power procurement by ev charging hubs.
\newblock In \emph{2023 IEEE Power \& Energy Society General Meeting (PESGM)}, pages 1--5. IEEE, 2023.

\bibitem[Paudel and Das(2023)]{paudel2023deep}
Diwas Paudel and Tapas~K Das.
\newblock A deep reinforcement learning approach for power management of battery-assisted fast-charging ev hubs participating in day-ahead and real-time electricity markets.
\newblock \emph{Energy}, 283:\penalty0 129097, 2023.

\bibitem[Subramanian and Das(2019)]{subramanian2019two}
Vignesh Subramanian and Tapas~K Das.
\newblock A two-layer model for dynamic pricing of electricity and optimal charging of electric vehicles under price spikes.
\newblock \emph{Energy}, 167:\penalty0 1266--1277, 2019.

\bibitem[DeForest et~al.(2018)DeForest, MacDonald, and Black]{deforest2018day}
Nicholas DeForest, Jason~S MacDonald, and Douglas~R Black.
\newblock Day ahead optimization of an electric vehicle fleet providing ancillary services in the los angeles air force base vehicle-to-grid demonstration.
\newblock \emph{Applied energy}, 210:\penalty0 987--1001, 2018.

\bibitem[Sarker et~al.(2015)Sarker, Dvorkin, and Ortega-Vazquez]{sarker2015optimal}
Mushfiqur~R Sarker, Yury Dvorkin, and Miguel~A Ortega-Vazquez.
\newblock Optimal participation of an electric vehicle aggregator in day-ahead energy and reserve markets.
\newblock \emph{IEEE transactions on power systems}, 31\penalty0 (5):\penalty0 3506--3515, 2015.

\bibitem[Kazemi et~al.(2017)Kazemi, Zareipour, Amjady, Rosehart, and Ehsan]{kazemi2017operation}
Mostafa Kazemi, Hamidreza Zareipour, Nima Amjady, William~D Rosehart, and Mehdi Ehsan.
\newblock Operation scheduling of battery storage systems in joint energy and ancillary services markets.
\newblock \emph{IEEE Transactions on Sustainable Energy}, 8\penalty0 (4):\penalty0 1726--1735, 2017.

\bibitem[Padmanabhan et~al.(2019)Padmanabhan, Ahmed, and Bhattacharya]{padmanabhan2019battery}
Nitin Padmanabhan, Mohamed Ahmed, and Kankar Bhattacharya.
\newblock Battery energy storage systems in energy and reserve markets.
\newblock \emph{IEEE Transactions on Power Systems}, 35\penalty0 (1):\penalty0 215--226, 2019.

\bibitem[Xu et~al.(2017)Xu, Wang, Dvorkin, Fern{\'a}ndez-Blanco, Silva-Monroy, Watson, and Kirschen]{xu2017scalable}
Bolun Xu, Yishen Wang, Yury Dvorkin, Ricardo Fern{\'a}ndez-Blanco, Cesar~A Silva-Monroy, Jean-Paul Watson, and Daniel~S Kirschen.
\newblock Scalable planning for energy storage in energy and reserve markets.
\newblock \emph{IEEE Transactions on Power systems}, 32\penalty0 (6):\penalty0 4515--4527, 2017.

\bibitem[Merten et~al.(2020)Merten, Olk, Schoeneberger, and Sauer]{merten2020bidding}
Michael Merten, Christopher Olk, Ilka Schoeneberger, and Dirk~Uwe Sauer.
\newblock Bidding strategy for battery storage systems in the secondary control reserve market.
\newblock \emph{Applied Energy}, 268:\penalty0 114951, 2020.

\bibitem[Nitsch et~al.(2021)Nitsch, Deissenroth-Uhrig, Schimeczek, and Bertsch]{nitsch2021economic}
Felix Nitsch, Marc Deissenroth-Uhrig, Christoph Schimeczek, and Valentin Bertsch.
\newblock Economic evaluation of battery storage systems bidding on day-ahead and automatic frequency restoration reserves markets.
\newblock \emph{Applied Energy}, 298:\penalty0 117267, 2021.

\bibitem[Hu et~al.(2022)Hu, Armada, and S{\'a}nchez]{hu2022potential}
Yu~Hu, Miguel Armada, and Mar{\'\i}a~Jes{\'u}s S{\'a}nchez.
\newblock Potential utilization of battery energy storage systems (bess) in the major european electricity markets.
\newblock \emph{Applied Energy}, 322:\penalty0 119512, 2022.

\bibitem[Ortega-Vazquez(2014)]{ortega2014optimal}
Miguel~A Ortega-Vazquez.
\newblock Optimal scheduling of electric vehicle charging and vehicle-to-grid services at household level including battery degradation and price uncertainty.
\newblock \emph{IET Generation, Transmission \& Distribution}, 8\penalty0 (6):\penalty0 1007--1016, 2014.

\bibitem[Mavrotas(2009)]{mavrotas2009effective}
George Mavrotas.
\newblock Effective implementation of the $\varepsilon$-constraint method in multi-objective mathematical programming problems.
\newblock \emph{Applied mathematics and computation}, 213\penalty0 (2):\penalty0 455--465, 2009.

\bibitem[Nash~Jr(1950)]{nash1950bargaining}
John~F Nash~Jr.
\newblock The bargaining problem.
\newblock \emph{Econometrica: Journal of the econometric society}, pages 155--162, 1950.

\bibitem[Charkhgard et~al.(2022)Charkhgard, Keshanian, Esmaeilbeigi, and Charkhgard]{charkhgard2022magic}
Hadi Charkhgard, Kimia Keshanian, Rasul Esmaeilbeigi, and Parisa Charkhgard.
\newblock The magic of nash social welfare in optimization: Do not sum, just multiply!
\newblock \emph{The ANZIAM Journal}, 64\penalty0 (2):\penalty0 119--134, 2022.

\bibitem[Khanal and Charkhgard()]{khanal4658986criterion}
Ashim Khanal and Hadi Charkhgard.
\newblock A criterion space search feasibility pump heuristic for solving maximum multiplicative programs.
\newblock \emph{Available at SSRN 4658986}.

\bibitem[Ben-Tal and Nemirovski(2001)]{ben2001polyhedral}
Aharon Ben-Tal and Arkadi Nemirovski.
\newblock On polyhedral approximations of the second-order cone.
\newblock \emph{Mathematics of Operations Research}, 26\penalty0 (2):\penalty0 193--205, 2001.

\bibitem[FDOT(2023)]{traffic_flow}
FDOT.
\newblock Traffic information.
\newblock \url{https://www.fdot.gov/statistics/trafficdata}, 2023.
\newblock (accessed : 03 March 2023).

\bibitem[Deng et~al.(2018)Deng, Tripathy, Tylavsky, Stowers, and Loehr]{deng2018demand}
Qian Deng, Sujit Tripathy, Daniel Tylavsky, Travis Stowers, and Jeff Loehr.
\newblock Demand modeling of a dc fast charging station.
\newblock In \emph{2018 North American Power Symposium (NAPS)}, pages 1--6. IEEE, 2018.

\bibitem[Narayan et~al.(2022)Narayan, Krishna, Misra, Vasan, and Sarangan]{narayan2022dynamic}
Ajay Narayan, Aakash Krishna, Prasant Misra, Arunchandar Vasan, and Venkatesh Sarangan.
\newblock A dynamic pricing system for electric vehicle charging management using reinforcement learning.
\newblock \emph{IEEE Intelligent Transportation Systems Magazine}, 14\penalty0 (6):\penalty0 122--134, 2022.

\bibitem[{Idaho National Laboratory}(2015)]{Idaho}
{Idaho National Laboratory}.
\newblock Plug-in electric vehicle and infrastructure analysis.
\newblock \url{https://inldigitallibrary.inl.gov/sites/sti/sti/6799570.pdf}, 2015.

\bibitem[ERCOT()]{ERCOT}
ERCOT.
\newblock Market information.
\newblock \url{https://www.ercot.com/mktinfo}.
\newblock (accessed : 28 September 2023).

\bibitem[Montgomery(2017)]{montgomery2017design}
Douglas~C Montgomery.
\newblock \emph{Design and analysis of experiments}.
\newblock John wiley \& sons, 2017.

\bibitem[Liu et~al.(2018)Liu, Xu, Yang, and Zhang]{liu2018distributionally}
Yongchao Liu, Huifu Xu, Shu-Jung~Sunny Yang, and Jin Zhang.
\newblock Distributionally robust equilibrium for continuous games: Nash and stackelberg models.
\newblock \emph{European Journal of Operational Research}, 265\penalty0 (2):\penalty0 631--643, 2018.

\bibitem[Peng et~al.(2021)Peng, Lisser, Singh, Gupta, and Balachandar]{peng2021games}
Shen Peng, Abdel Lisser, Vikas~Vikram Singh, Nalin Gupta, and Eshan Balachandar.
\newblock Games with distributionally robust joint chance constraints.
\newblock \emph{Optimization Letters}, pages 1--23, 2021.

\end{thebibliography}






\end{document}